\documentclass[epj]{svjour}
\usepackage{graphics}
\usepackage{amsmath} 
\usepackage{amssymb}
\usepackage{graphicx}
\begin{document}
\title{Beta-delayed proton emission from $^{21}$Mg}
\author{M. V. Lund\inst{1}\thanks{e-mail: mvl07@phys.au.dk}, M. J. G. Borge\inst{2, 5}, J. A. Briz\inst{2}, J. Cederk\"{a}ll\inst{3}, H. O. U. Fynbo\inst{1}, J. H. Jensen\inst{1}, B. Jonson\inst{4}, K. L. Laursen\inst{1}, T. Nilsson\inst{4}, A. Perea\inst{2}, V. Pesudo\inst{2}, K. Riisager\inst{1}, \and O. Tengblad\inst{2}}                     
%
%
\institute{Department of Physics and Astronomy, Aarhus University, DK-8000 Aarhus C, Denmark \and 
Instituto de Estructura de la Materia, CSIC, E-28006 Madrid, Spain \and 
Department of Nuclear Physics, Lund University, SE-221 00 Lund, Sweden \and 
Department of Fundamental Physics, Chalmers University of Technology, SE-412 96 G\"{o}teborg, Sweden \and
ISOLDE, PH Department, CERN, CH-1211 Geneva 23, Switzerland}

\date{Received: date / Revised version: date}
%
\abstract{Beta-delayed proton emission from $^{21}$Mg has been measured at ISOLDE, CERN, with a detection setup including particle identification capabilities. $\beta$-delayed protons with center of mass energies between 0.39$\,$MeV and 7.2$\,$MeV were measured and used to determine the half life of $^{21}$Mg as $118.6\pm 0.5\,$ms. From a line shape fit of the $\beta p$ branches we extract spectroscopic information about the resonances of $^{21}$Na. Finally an improved interpretation of the decay scheme in accordance with the results obtained in reaction studies is presented.
} 

\authorrunning{M. V. Lund et al.}
\titlerunning{$^{21}$Mg $\beta$-decay}
\maketitle

\section{Introduction}
\label{intro}
The decay of drip-line nuclei is characterized by the large available $\beta$-decay energy which together with the small particle separation energy are responsible for the many open decay channels. At the proton drip-line we mainly observe $\beta$-delayed single particle emission whereas beta-delayed multi-particle emission modes are more rare \cite{bib:Borge08,bib:Pfutzner,bib:Borge13}. The study of such nuclear decay modes allows us to obtain detailed information about the level structure of the daughter nuclei and to characterize the $\beta$-decay strength distribution. From studies of the energy spectrum of the $\beta$-delayed particles it is also possible to gain valuable spectroscopic information about the particle emitting resonances like partial and total decay widths, spin, and parity - all from a detailed description of the line shape of the emitted particles.

In the case of $^{21}$Mg, $Q_{EC}=13.098(16)\,$MeV \cite{bib:Nubase}, the $\beta^+$-decay to bound states of $^{21}$Na has a branching ratio of 67(7)$\,$\% \cite{bib:Sextro}. The only other identified decay mode is $\beta$-delayed proton emission to states in $^{20}$Ne, but $\beta\alpha$, $\beta\alpha p$, and $\beta p\alpha$ are all energetically allowed. As the separation energies for such decays are quite high, $S_\alpha = 6561.3(4)\,$keV and $S_{p\alpha} = S_{\alpha p} = 7161.5(3)\,$keV \cite{bib:Nubase}, it is most likely to observe such low intensity decay modes as emissions from the Isobaric Analog State, IAS, with $T_z=\frac{3}{2}$ at an excitation energy of 8975(4)$\,$keV, as this state is strongly fed in the $\beta$-decay due to the similarity of the structure with the ground state of $^{21}$Mg and the position of the IAS inside the $Q_{EC}$-window.

An earlier experiment by Sextro et al. \cite{bib:Sextro} in 1973 used different $\Delta E$-$E$ Si telescope combinations to cover the entire energy range of the $Q_{EC}$ energy window, and a helium-jet transport system to get the $^{21}$Mg activity away from the driver beam. The energy resolution obtained depended on the detector combination used but ranged between 25$\,$keV and 45$\,$keV FWHM. With this setup they reported 25 different $\beta$p decay branches and placed upper limits on the intensity of the $\beta\alpha$ decay branch between the IAS of $^{21}$Na and the ground state of $^{17}$F. A more recent experiment \cite{bib:JCThomas} also studied the $\beta$-decay of $^{21}$Mg but they did not observe all the $\beta$p decay branches of \cite{bib:Sextro}. However, they observed four $\beta p\gamma$ transitions. The proton intensities measured by \cite{bib:Sextro} and \cite{bib:JCThomas} are not consistent.

In Sect. \ref{sec:Experiment} we will describe the experiment including the beam production, the detection setup, and the energy calibrations. In Sect. \ref{anal} we go through the analysis. First we present the measured charged particle spectra, then we present an improved value for the half-life followed by a time distribution analysis. Finally we describe the line shape fit of the charged particles. In Sect. \ref{results} we present an improved interpretation of the decay scheme based on the current measurement in combination with recent $^{20}$Ne(p,p) scattering experiments.

\section{Experiment}
\label{sec:Experiment}
The experiment was performed at the ISOLDE facility \cite{bib:isol} at CERN, Switzerland.

\subsection{Beam production}
The production of the exotic $^{21}$Mg beam was achieved by using a SiC target bombarded with $1.4\,$GeV protons, and the laser ion-source RILIS \cite{bib:Rilis} for isotope specific ionization of Mg. The 60$\,$keV radioactive ion beam was guided through the High Resolution Separator (HRS) \cite{bib:isol} to separate the desired Mg isotope from the isobaric background of Na. However, a significant Na contamination remained in the delivered beam due to the three orders of magnitude larger production yield of Na over Mg and due to the closeness of the masses, $\Delta M=13.098(16)\,$MeV - the resolution of the HRS is $M/\Delta M=5000$. This contamination of Na was directly observed when the RILIS ionization of Mg was turned off - under such conditions no Mg was observed as expected. To further suppress the Na contamination we made use of the fact that the proton beam on the ISOLDE target is pulsed with pulse spacing a multiple of 1.2$\,$s. This, together with the fact that $^{21}$Mg and $^{21}$Na have a large difference in half-lives, 122(2)$\,$ms and 22.49(4)$\,$s respectively \cite{bib:Nubase}, and that the timescale for Mg ions to diffuse out of the target, be ionized, and transported to the setup is of the order of 100$\,$ms, provides a natural way of suppressing the isobaric contamination of Na by only letting the beam into the setup during the first 300$\,$ms following proton impact on target. From measurements on mass 20, which show a similar difference in the half-lives of $^{20}$Mg and $^{20}$Na, a ratio of $\frac{^{20}\text{Na}}{^{20}\text{Mg}}=40$ was found - we assume the ratio to be of similar magnitude on mass 21. An average of $390$ $^{21}$Mg ions per $\mu$C of proton beam on target were delivered to the detection system over a total beam time of $\sim3.5$ hours. The average proton current delivered on the SiC production target was $\sim1.9\,\mu$A resulting in approximately $9\cdot 10^6$ $^{21}$Mg ions delivered to the detection chamber.

\begin{figure}[h!]
\resizebox{0.50\textwidth}{!}{
\includegraphics{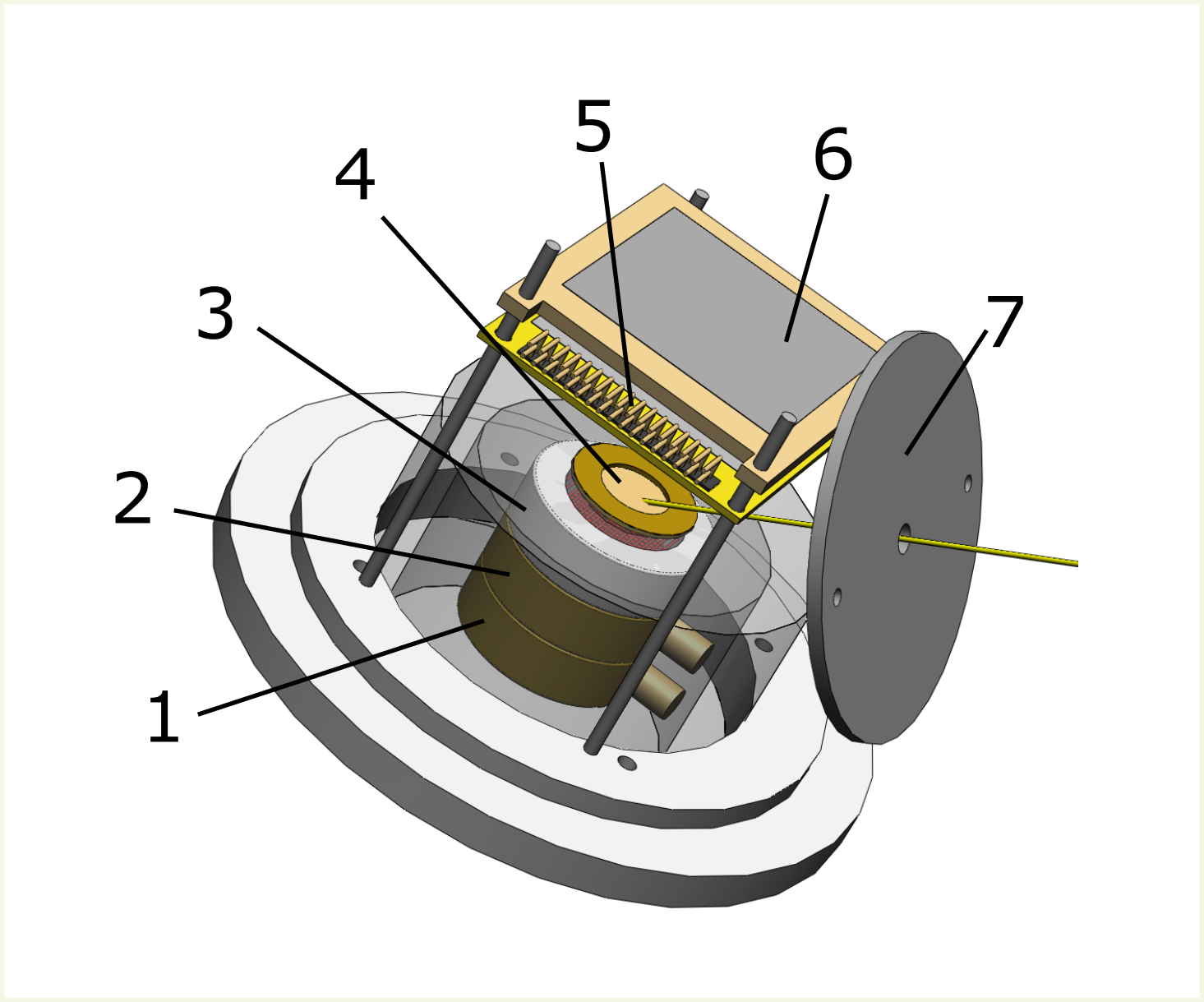}
}
\caption{(Color online) Drawing of the charged particle detection setup. The beam is coming from the right, passing through the collimator (7) and is implanted in the polypropylene window (4) of the gas detector (3). Behind the gas detector, sitting inside the gas volume, are placed two silicon detectors (1 and 2), which in total form the $\Delta E$-$E$-$E$ charged particle telescope. Opposite to the Gas-Si-Si telescope is the Si-Si telescope with the DSSSD as the front detector (5) and backed by a thick silicon pad detector (6).}
\label{fig:setup}
\end{figure}

\subsection{Detection system}
The experimental setup used for the measurement of the $^{21}$Mg $\beta$-decay can be seen on Figure \ref{fig:setup}. It was designed to identify different low-energy, light, charged particles by the use of a Gas-Si-Si charged particle telescope with an opposing Si-Si telescope. The $\beta$-particle response of the gas detector is very small and it allowed effective discrimination between $\beta$-particles, protons, and $\alpha$-particles. The beam of $^{21}$Mg was implanted in the $1\,\mu$m thick polypropylene window (no. 4 in Figure \ref{fig:setup}) confining the gas in the Gas-Si-Si telescope. The gas detector was backed by a 300$\,\mu$m thick silicon detector with a solid angle coverage of $7.1(9)\,\%$. It was further backed by a 500$\,\mu$m thick silicon detector with a solid angle coverage of $3.6(1)\,\%$. Both silicon detectors were located inside the gas volume. The three detectors were circular with an area of 300$\,$mm$^2$.

The opposing Si-Si charged particle telescope consisted of a 61$\,\mu$m thick, $5\times 5\,$cm$^2$ Double Sided Silicon Strip Detector (DSSSD) with a solid angle coverage of $7.7(3)\,\%$. It was backed by a 1000$\,\mu$m thick silicon pad detector of the same area and with a solid angle coverage of $6.6(4)\,\%$. This telescope is ideal for measuring the intensities of the $\beta$-delayed charged particles as systematic effects such as $\beta$-summing \cite{bib:BetaSumming} are minimized due to the small solid angle of the individual pixels, $\Omega_{\text{pixel}}=0.030(1)\,\%$. In addition, the two charged particle telescopes will in combination allow us to set coincidence or anti-coincidence gates which is a powerful tool in the identification of low-energy and low intensity branches. 

\subsection{Energy calibration}
\label{sec:EnergyCalib}
The geometry and energy calibrations of the DSSSD and the first silicon detector in the Gas-Si-Si telescope were made with the $^{21}$Mg beam itself, as several strong $\beta$-delayed proton branches are available and their energy is accurately known from reaction studies. For the first silicon detector in the Gas-Si-Si telescope we used the protons with center-of-mass energies 1862(2)$\,$keV, 2036(5)$\,$keV, and 4904(4)$\,$keV \cite{bib:Sextro} - the three most intense proton lines. For the DSSSD we used the protons with center-of-mass energies 1320(10)$\,$keV, 1862(2)$\,$keV, and 2036(5)$\,$keV \cite{bib:Sextro}, which are the three most intense proton lines that are stopped in this detector. The measured energy spectra can be seen in Figure \ref{fig:ProtonSpectrumSi1} and Figure \ref{fig:ProtonSpectrumDSSSD}.

For the energy calibration of the thick silicon pad detector backing the DSSSD a quadruple $\alpha$-source was used ($^{148}$Gd, $^{239}$Pu, $^{241}$Am, $^{244}$Cm). As $\alpha$-particles and protons have different stopping powers, the energy calibration will not accurately reproduce the proton energies. However, the stopping power can be divided into an ionizing part and a non-ionizing part. Both parts of the stopping power will differ for protons and $\alpha$-particles and this effect can be corrected for as described in Ref. \cite{bib:Lennard}. Accordingly, the energy calibration can be transformed by first correcting the energy deposited in the detector for the difference in the non-ionizing energy lost for the two types of particles. This is as an average over energy given as $(\Delta E_n)_{^1\text{H}}=1\,$keV and $(\Delta E_n)_{^4\text{He}}=9\,$keV. Secondly, a correction for the difference in the ionizing energy lost must be made. The ionizing part of the energy lost by an $\alpha$-particle can be correcting with the factor $C(^1$H$)/C(^4$He$)=0.986(2)$, where $C$ is the slope of a linear fit to the pulse height versus the deposited energy. 

The resulting linear energy calibrations give a FWHM energy resolution of 39.3(2)$\,$keV for the DSSSD, 53.9(3)$\,$keV for the sum spectrum of the Si-Si telescope detectors, and 49.0(3)$\,$keV for the first silicon detector of the Gas-Si-Si telescope. These values are obtained using the line shape described in Sect.~\ref{sec:LineShape}, Eq. (\ref{eq:LineShape}).

\section{Analysis}
\label{anal}
In Sect.~\ref{sec:Spectra} the measured spectra are presented with a focus on the new decay branches. In Sect. \ref{sec:HalfLife} the determination of the half-life of $^{21}$Mg is explained. Any new decay branches in the decay of $^{21}$Mg could in principle be explained as isobaric contamination in the beam coming mainly from $^{21}$Na as mentioned in Sect.~\ref{sec:Experiment}. To rule this out a goodness-of-fit test of the time distributions of the new decay branches is performed in which the time distributions are compared with the $^{21}$Mg time distribution, this is presented in more detail in Sect.~\ref{sec:TimeTest}. Finally in Sect.~\ref{sec:LineShape} a detailed analysis of the proton line shape is described from which spectroscopic information about the daughter nuclei is extracted.

\begin{figure}
\resizebox{0.50\textwidth}{!}{
  \includegraphics{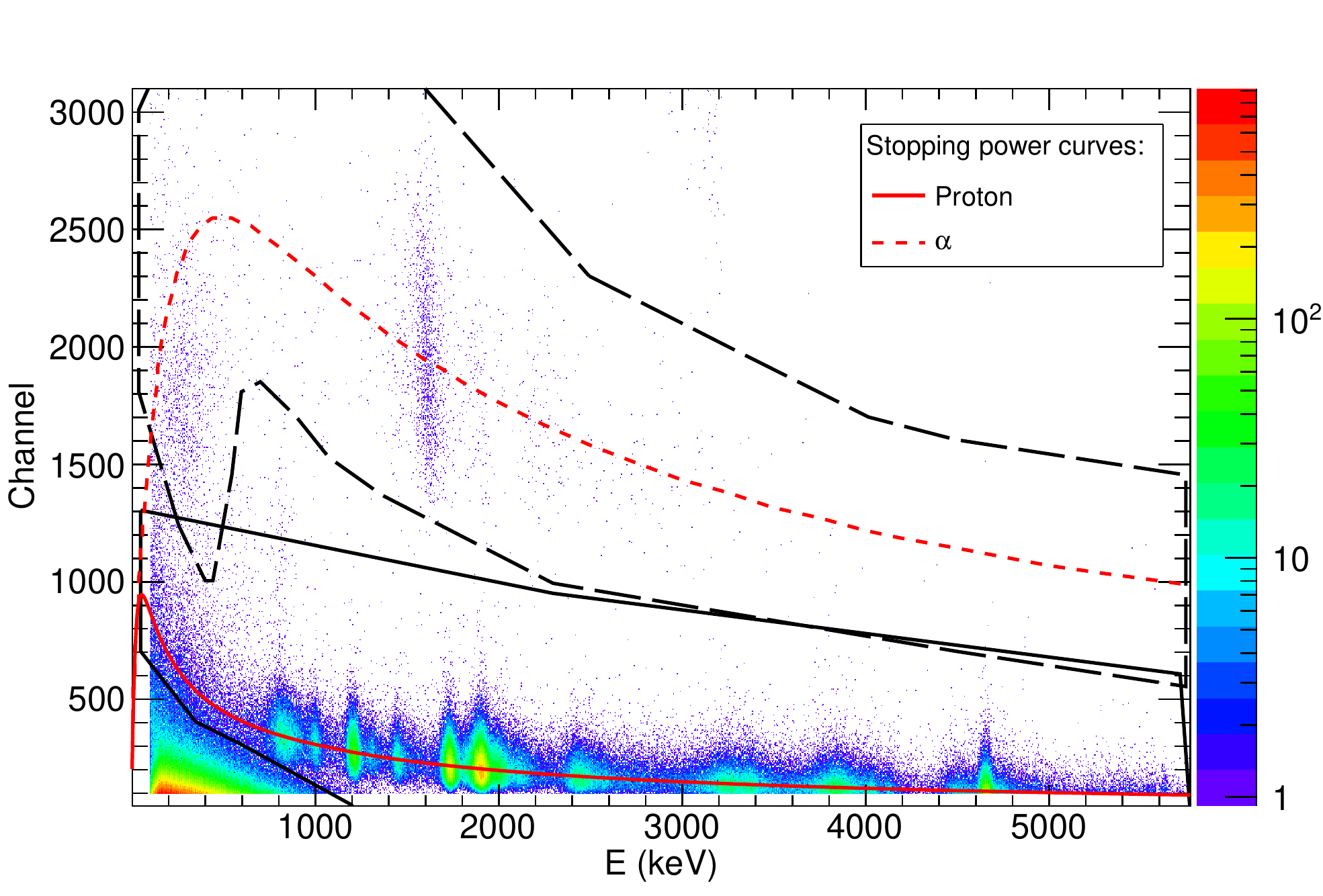}
}
\caption{(Color online) $\Delta E$-$E$ plot from the Gas-Si-Si telescope with the gas channel on the vertical axis and the deposited energy in the first silicon detector on the horizontal axis. The scaled stopping powers for $\alpha$-particles and protons in silicon is shown on top of the data in dashed and solid red respectively. The graphical cut used for the $\alpha$-particles is shown with the long dashed black closed line and the cut for the protons is shown by the solid black closed line. The events located at 3.18$\,$MeV in the silicon detector and above channel 2300 in the gas detector is contamination from a source of $^{148}$Gd.}
\label{fig:BananaPlot}
\end{figure}

\subsection{Spectra}
\label{sec:Spectra}
The data from the Gas-Si-Si charged particle telescope are presented as a $\Delta E$-$E$ spectrum in Figure \ref{fig:BananaPlot}. Stopping power curves \cite{bib:SRIM} for $\alpha$-particles and protons in silicon are drawn in the figure. The stopping powers are rescaled in order to represent the total energy loss in the collection foil, the gas detector, and the silicon dead layer. The data match the stopping power curves making it clear that both protons and $\alpha$-particles are present in the collected data sample. This reveals the presence of the two decay modes $\beta p$ and $\beta\alpha$. A more detailed analysis of the $\beta\alpha$ decay branches can be found in Ref. \cite{bib:Lund} together with a discussion of the first observation of the rare $\beta$p$\alpha$ decay mode.

\begin{figure*}
\resizebox{1.00\textwidth}{!}{
\includegraphics{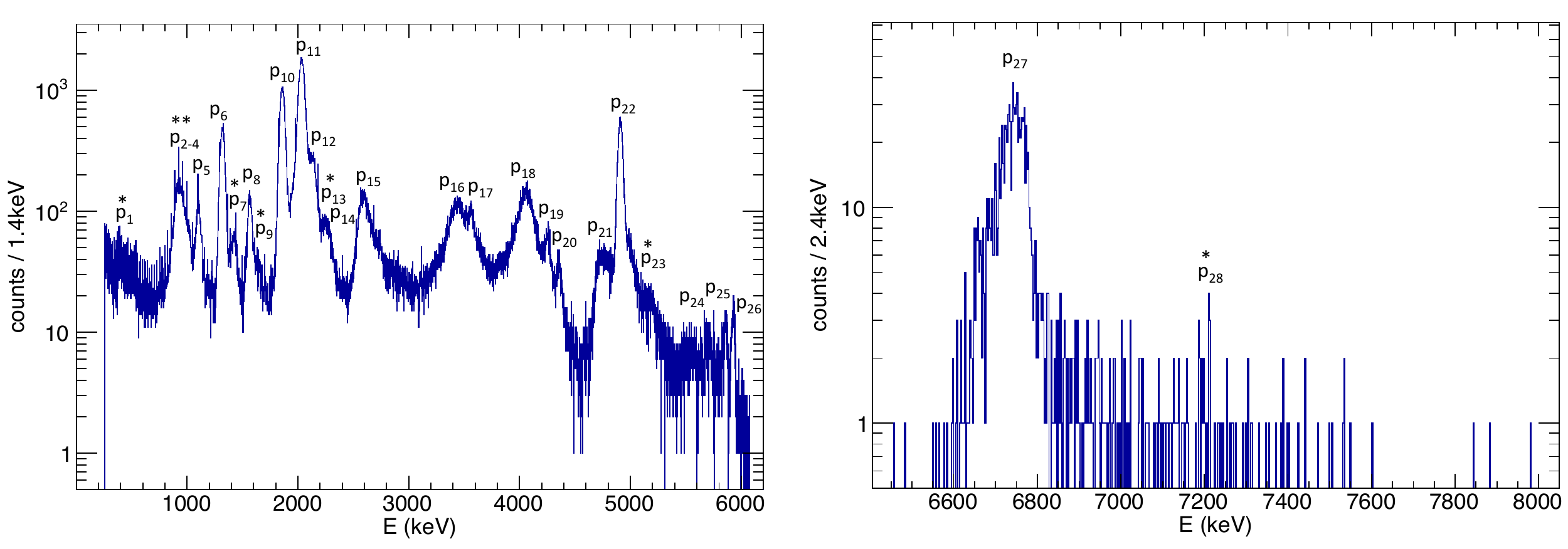}
}
\caption{(Color online) Proton singles spectrum in the Gas-Si-Si telescope with the proton center-of-mass energy shown. New proton branches are marked with a $\star$. A $\star\star$ indicates two new proton branches.  \textit{Left:} the first silicon detector (number 2 in Figure \ref{fig:setup}) of the Gas-Si-Si telescope. \textit{Right:} the second silicon detector (number 1 in Figure \ref{fig:setup}) of the Gas-Si-Si telescope. Note that the energy calibration of the second silicon detector suffer from systematics.}
\label{fig:ProtonSpectrumSi1}
\end{figure*}

\begin{figure*}
\resizebox{1.00\textwidth}{!}{
\includegraphics{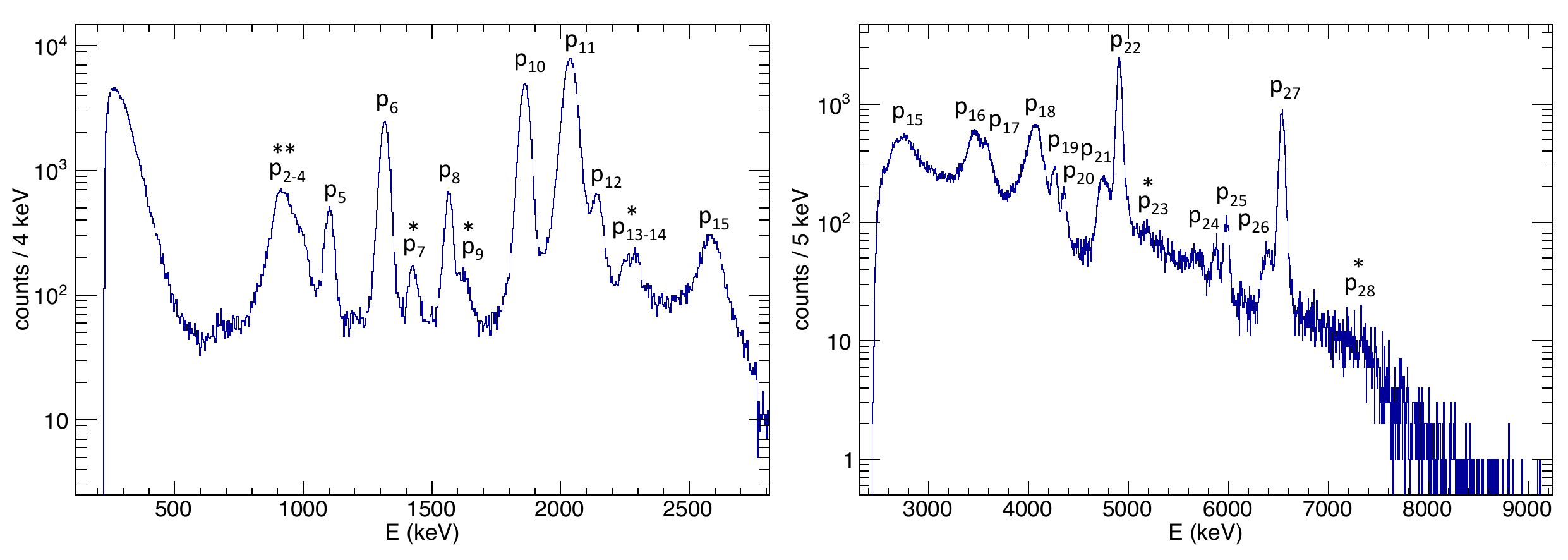}
}
\caption{(Color online) Proton singles spectrum in the Si-Si telescope with the proton center-of-mass energy shown. New proton branches are marked with a $\star$. A $\star\star$ indicates two new proton branches. \textit{Left:} The proton energy, $E_{\text{cm}}$, as measured in the DSSSD (number 5 in Figure \ref{fig:setup}). \textit{Right:} The proton energy, $E_{\text{cm}}$, as measured by the sum of the DSSSD and the backing silicon pad detector (number 6 in Figure \ref{fig:setup}). Note that p$_{15}$ punch-through the center part of the DSSSD but is stopped in the outer parts of the detector due to the larger angle of incidence. The proton branch p$_{28}$ is placed tentatively at the expected position in the energy spectrum.}
\label{fig:ProtonSpectrumDSSSD}
\end{figure*}

To get a clean proton spectrum from the Gas-Si-Si telescope two gates have been applied to the data. First, a gate to remove the response of the high energy protons that punch through the first silicon detector is applied. Second, a condition to reduce the $\beta$-particle response and the recoils is applied. The same gate also reduces the number of $\alpha$-particles in the spectrum. The condition is shown as the solid black closed line in Figure \ref{fig:BananaPlot}. The resulting proton spectrum from the first silicon detector can be seen in Figure \ref{fig:ProtonSpectrumSi1}. It is obvious that the gated proton spectrum has almost no background except below approximately 1150$\,$keV, where the $\beta$-particle contribution cannot be completely removed.

While the energy resolution achieved in this experiment is slightly worse than those of the earlier experiment \cite{bib:Sextro}, more statistics is collected, which makes it possible to see low intensity transitions not observed before. In general, the proton spectrum exhibits the same features as observed in the previous measurements, however, a closer look reveals several proton branches not previously observed with $E_{\text{cm}}$(p$_1$)$=396(3)\,$keV,  $E_{\text{cm}}$(p$_7$) = 1427.1(9)$\,$keV, $E_{\text{cm}}$(p$_9$) = 1630.0(15)$\,$keV, $E_{\text{cm}}$(p$_{23}$)$=5171(7)\,$keV, and $E_{\text{cm}}$(p$_{28}$) = 7.2(3)$\,$MeV. The new proton line p$_{28}$ is observed in the singles proton spectrum of the second silicon detector in the Gas-Si-Si telescope (see Figure \ref{fig:ProtonSpectrumSi1}) with 22(9) events after background subtraction. 

A new interpretation of the line shape of the peak p$_{2-4}$ gives rise to three $\beta p$ transitions with energies: $E_{\text{cm}}$(p$_2$)$=906.2(14)\,$keV, $E_{\text{cm}}$(p$_3$)$=919(18)\,$keV, and $E_{\text{cm}}$(p$_4$)=936.8 $(13)\,$keV. In \cite{bib:Sextro} only one $\beta p$ transition was assumed here. More details can be found in Sect. \ref{sec:Interpretation}.

In the Si-Si telescope similar gates have been set to reduce the $\beta$-particle response at low energies and to remove punch-through and $\beta$ particles in the DSSSD. The resulting proton spectra can be seen in Figure \ref{fig:ProtonSpectrumDSSSD} and they show the same main features as in the Gas-Si-Si telescope except in the case of p$_1$, which is hidden by the more pronounced $\beta$-particle response of the DSSSD compared to the gas detector.

Due to the better energy resolution in the DSSSD it appears that the proton line just above p$_{12}$, which in the Gas-Si-Si telescope looks like a single broad proton line, actually consist of two close lying proton lines, p$_{13}$ and p$_{14}$. Their energy is measured to be $E_{\text{cm}}$(p$_{13}$) = 2263(4)$\,$keV and $E_{\text{cm}}$(p$_{14}$) = 2302(2)$\,$keV. The proton line p$_{23}$ is also more pronounced in the Si-Si telescope than in the Gas-Si-Si telescope.

\subsection{Half-life determination}
\label{sec:HalfLife}
\begin{figure}
\resizebox{0.50\textwidth}{!}{
  \includegraphics{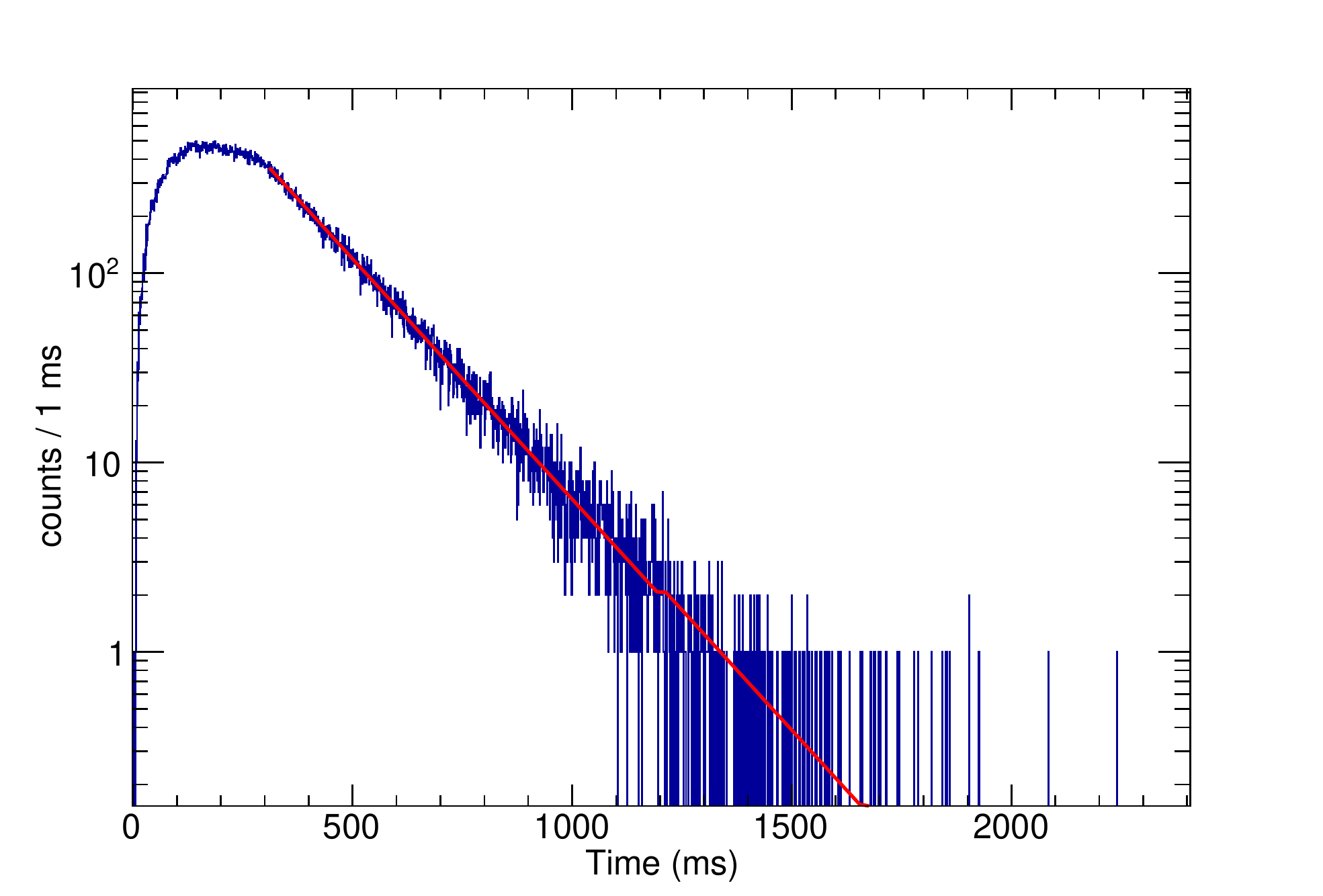}
}
\caption{(Color online) Half-life fit for the $^{21}$Mg decay. Fitting is performed between 300--2400$\,$ms. The lower limit is the time when the beam is no longer allowed into the detection setup and the upper limit is chosen to coincide with two times the ISOLDE target proton beam pulse distance of 1.2$\,$s. At this upper limit in time there is no activity left.}
\label{fig:HalfLife}
\end{figure}

The half-life of $^{21}$Mg was determined from the time distribution of the two most intense proton branches in the DSSSD, p$_{10}$ and p$_{11}$. The time distribution, see Figure \ref{fig:HalfLife}, is fitted with a function describing the standard radioactive decay law $f(x) = p_0\cdot p_1\cdot \text{e}^{-p_0\cdot x}$ and the fitting parameters are minimized by the use of the \texttt{MINOS} error estimation technique in the \texttt{MINUIT2} minimization package. A standard Poisson log-likelihood method was used in order to include bins with zero counts and to obtain a more reliable fit when small count numbers are present. The resulting half-life is $T_{1/2}=118.6(5)\,$ms with a $\chi^2/$ndf $ =1234/2099=0.59$ which based on the result of \cite{bib:Bergmann} is a good fit. The half-life obtained here and the previous evaluation of the half-life, $T_{1/2}=122(2)\,$ms \cite{bib:Nubase}, are within two standard deviations of each other. The uncertainty on our new determination is a factor of four lower. The new value of $T_{1/2}=118.6(5)\,$ms will be used in the calculation of the log$(ft)$-values in Sect. \ref{sec:logft}.

\subsection{Time distribution analysis}
\label{sec:TimeTest}
As mentioned earlier the implanted ion beam is contaminated with $^{21}$Na, but other less produced contaminants could also be present. As the intensity of several of the new $\beta$p branches is low, an additional cross-check of the assignment is desirable. This will be provided by analysing the time distribution of the events.

The large difference in halflives of $^{21}$Mg and $^{21}$Na, 118.6(5)$\,$ms and 22.49(4)$\,$s \cite{bib:Nubase}, and the fact that the proton beam on the ISOLDE target is pulsed with pulse spacing a multiple of 1200$\,$ms, provide a natural way of distinguishing between the two decays. The timescale for Mg ions to diffuse out of the target, be ionized and transported to the setup is of the order of 100$\,$ms, and the beam was let into the setup during the first 300$\,$ms following proton impact on target. The resulting time distribution of the two decays is therefore somewhat complex. Since some of the $\beta$p branches have quite low statistics it is better to compare their time distribution to a reference distribution rather than attempting a fit.  To do this as efficiently as possible a goodness-of-fit test based on empirical distribution function (EDF) statistics is employed. We note that this test will also be able to discriminate against contaminants from other activities appearing in the ion beam.

The EDF statistics \cite{bib:Ste86} are known to give more powerful goodness-of-fit tests than e.g.\ the often employed $\chi^2$ test. They compare the shape of the data sample with a reference shape by measuring the distance between the two cumulated distributions. For the case of binned data the experimental and reference distributions have values $EDF_i$ and $F_i$ in bin $i$, and the most frequently used EDF statistics are \cite{bib:Cho94} Kolmogorov-Smirnov
\begin{equation*}
D = \sqrt{N} max_i |EDF_i-F_i|,
\end{equation*}
 Cramer-Von Mises 
 \begin{equation*}
 W^2 = N \sum_i (EDF_i-F_i)^2 p_i,
 \end{equation*}
 and Anderson-Darling 
 \begin{equation*}
 A^2 = N \sum_i \frac{(EDF_i-F_i)^2 p_i}{F_i(1-F_i)}, 
 \end{equation*}
 where $N$ is the total number of counts and $p_i$ is the probability to be in bin $i$ in the reference distribution. The values of these statistics corresponding to specific confidence levels must in the general case (where the reference distribution contains parameters and is fitted to the data) be found by simulations. In our case the reference distribution is taken to be the time distribution recorded for events within the proton gate in Figure \ref{fig:BananaPlot} and with energy above 1150$\,$keV. This region is expected to contain only protons from the decay of $^{21}$Mg. Since this is a fully specified distribution, confidence levels should be close to the ones for a standard uniform distribution. We have carried out Monte Carlo simulations to evaluate the confidence levels with 40000 randomly generated spectra each with 400 counts (varying the number of counts does not change the results much) and the resulting confidence levels are given in Table \ref{tab:Significance}. They are very close to the ones found for $W^2$ and $A^2$ for a uniform binned distribution in \cite{bib:Cho94} and also close to the values for unbinned data, whereas our results for $D$ are 0.05--0.15 lower than for the unbinned case.

\begin{table}
\caption{Established confidence levels for the three different statistical tests of the time distributions based on Monte Carlo simulations. The confidence levels are calculated for the hypotheses that the time distribution in question belong to the decay of $^{21}$Mg. Hence a good fit will have a confidence level less than 95$\,\%$.}
\label{tab:Significance} 
\centering
\begin{tabular}{ccccccc}
\hline\noalign{\smallskip}
c.l.: & 75$\,\%$ & 85$\,\%$ & 90$\,\%$ & 95$\,\%$ & 97.5$\,\%$ & 99$\,\%$ \\
\noalign{\smallskip}\hline\noalign{\smallskip}
D & 0.98 & 1.10 & 1.18 & 1.31 & 1.43 & 1.57 \\
W$^2$ & 0.21 & 0.29 & 0.35 & 0.46 & 0.57 & 0.72 \\
A$^2$ & 1.24 & 1.63 & 1.94 & 2.49 & 3.09 & 3.85 \\
\noalign{\smallskip}\hline
\end{tabular}
\end{table}

\begin{table}
\caption{Results of the three different goodness-of-fit tests of the time distribution to settle if the events in question do belong to the $^{21}$Mg decay. To be compared with the confidence levels quoted in Table \ref{tab:Significance}.}
\label{tab:TimeResult} 
\centering
\begin{tabular}{ccccccccc}
\hline\noalign{\smallskip}
 & p$_3$ & p$_7$ & p$_9$ & p$_{28}$ \\
\noalign{\smallskip}\hline\noalign{\smallskip}
D & 1.22 & 0.95 & 1.03 & 0.78 \\
W$^2$ & 0.33 & 0.15 & 0.21 & 0.13 \\
A$^2$ & 1.46 & 1.01 & 1.01 & 0.86 \\
\noalign{\smallskip}\hline
\end{tabular}
\end{table}

The described goodness-of-fit tests were applied to all the new decay branches and the results of the tests can be seen in Table \ref{tab:TimeResult} except for the low energy proton branch p$_1$, which will be discussed below. For the proton branches p$_7$, p$_9$, and p$_{28}$ the goodness-of-fit tests return a confidence level lower than $85\,\%$, i.e. the proton branches are consistent with coming from the $^{21}$Mg decay. The result for p$_3$ varies among the three tests with the result of the Kolmogorov-Smirnov test being only marginally consistent with p$_3$ belonging to the decay of $^{21}$Mg. However, the result of the Cramer-Von Mises test shows a confidence level smaller than 90$\,$\% and the result of the Anderson-Darling test gives a confidence level smaller than 85$\,$\% which means that p$_3$ is also consistent with belonging to the $^{21}$Mg decay.

\begin{figure}
\resizebox{0.50\textwidth}{!}{
\includegraphics{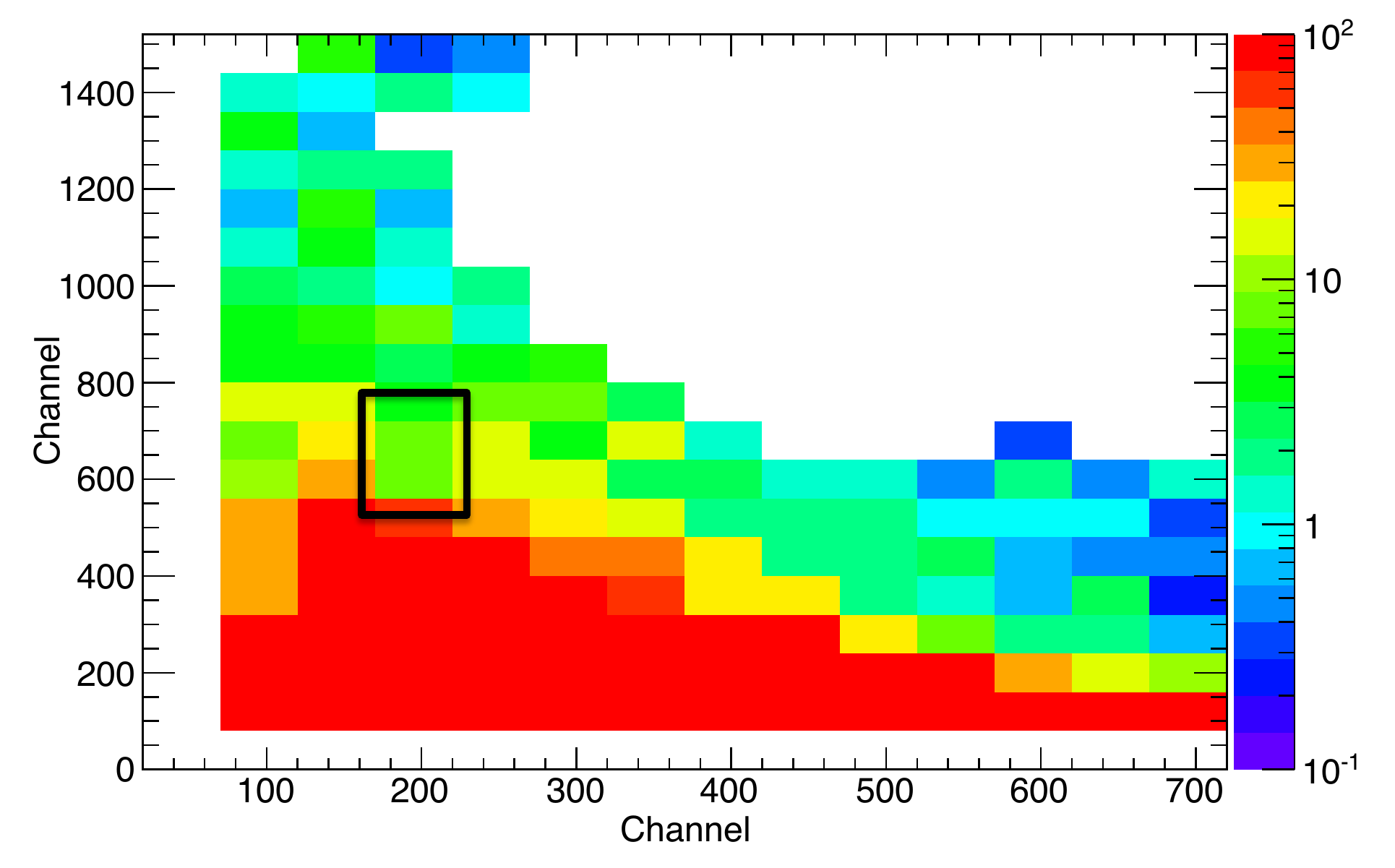}
}
\caption{(Color online) Result of the Anderson-Darling goodness-of-fit test, $A^2$, of the $\beta$-particle region in the $\Delta E$-$E$ plot with the $^{21}$Mg($\beta$p) time distribution as the reference distribution. The vertical axis shows the channel number of the gas detector and the horizontal axis shows the channel number of the first silicon detector in the Gas-Si-Si telescope. The proton branch p$_1$ is marked with a black rectangle.}
\label{fig:MixedRegionTimeTest}
\end{figure}

The proton branch p$_1$ is located in an energy region of Figure \ref{fig:BananaPlot} which contains both $\beta$-particles from Na and Mg but also recoils. It means that we do not expect the goodness-of-fit test to confirm p$_1$ as part of the $^{21}$Mg decay, which is exactly what we observe. We thus confirm the reliability of the method. However, in order to support the assignment of p$_1$ we applied the Anderson-Darling goodness-of-fit test, $A^2$, to the low energy region of the $\Delta E$-$E$ spectrum after having divided the region into small rectangles. The time distribution of the events in each rectangle was compared with the $^{21}$Mg time distribution and the value of $A^2$ for each rectangle can be seen in Figure \ref{fig:MixedRegionTimeTest}. For the time distribution of the individual rectangles to have a confidence level smaller than 95$\,$\% a value of $A^2<2.49$ is needed according to Table \ref{tab:Significance}. The interesting proton branch p$_1$ is marked with the black rectangle. Only for the events here does the goodness-of-fit test show a better agreement with the $^{21}$Mg time distribution than in adjacent rectangles - at lower and larger silicon channel numbers $A^2$ is significantly larger. It tells us that here we have a signal which could belong to a different time distribution than the general background in the region. As the position also coincides with the proton stopping power in silicon as shown in Figure \ref{fig:BananaPlot} we conclude that it is very likely to be a low energy $\beta p$ decay branch from $^{21}$Mg with a large background of $\beta$-particles from $^{21}$Na.

\subsection{Line shape fit for charged particles}
\label{sec:LineShape}
\subsubsection{Line shape}
\label{subsec:LineShape}
As a first order approximation of the line shape local fits with a normalized Gaussian distribution and a constant background were used to extract the number of protons measured. This is the same analysis method as used in \cite{bib:Sextro}. However, this method turns out to give inconsistent results when comparing the Gas-Si-Si telescope and the Si-Si telescope results. Hence, a more physically correct line shape was developed to better account for the physics case.

First, an approximation on the resonant behavior of the decay probability for $\beta$-delayed emission of charged particles is made. As this is a very complex quantum mechanical problem it is best solved with the R-matrix theory, \cite{bib:Rmatrix}. However, we will use an approximation of the R-matrix theory by assuming non-interfering and narrow resonances, which gives a simple modeling of the decay probability as a Breit-Wigner distribution. The signature of interference and of broad resonances are an asymmetric line shape of the emitted charged particles which we will take into consideration when needed when fitting the observed spectrum.

Second, the detector response is approximated as a normalized and integrated Gaussian distribution. This describes the number of counts per channel in the detector from a monoenergetic beam and is a sufficient approximation in this case. 

Third, the recoil broadening is not included in the final line shape. The reason is that the detector response broadening is larger than the recoil broadening and it will thus only give rise to a small perturbation on the determined widths. The maximum recoil shift $T_{\text{max}}$ will be between 20--26$\,$keV based on the expression given in \cite{bib:Bha02} with the exact value depending on the excitation energy in $^{21}$Na. Thus the FWHM of the recoil broadening, which will be less than $2\cdot T_{\text{max}}$, is smaller than the best energy resolution discussed in Sect. \ref{sec:EnergyCalib}.

The final function used for extracting the number of events in each peak, assuming a background level of zero, is
\begin{align}
\psi (E) = &\frac{A}{2}\left[ Erf\left(\frac{E+\Delta-E_0}{\sqrt{2}\sigma}\right) - Erf\left(\frac{E-\Delta-E_0}{\sqrt{2}\sigma}\right)\right] \nonumber \\ 
& \nonumber \\ 
 &\times \frac{\Gamma}{2\pi}\frac{1}{(E-E_0)^2+(\Gamma/2)^2}\, ,
 \label{eq:LineShape}
\end{align}
where $Erf(x)=\frac{2}{\sqrt\pi}\int_0^xe^{-u^2}du$. The parameter $A$ is the number of events in the peak, $E_0$ is the centroid energy of the peak, $\sigma$ is the Gaussian width of the detector response function, which is the same as the detector resolution. The parameter $\Delta$ is half the bin-width of the spectrum, which is a fixed property of the individual spectra. Finally, $\Gamma$ is the Breit-Wigner width of the resonance emitting the corresponding charged particle. The convolution is performed numerically.

\subsubsection{Fitting procedure}
The fitting procedure adopted in the analysis is to first determine the Gaussian width $\sigma$ of the detector response by fitting a well known and narrow proton peak. For the DSSSD of the Si-Si telescope a fit to the proton line p$_{10}$, emitted from the $4294.3(6)\,$keV resonance in $^{21}$Na, with the Breit-Wigner width fixed to the literature value $\Gamma_{\text{tot}}=3.93(10)\,$keV gives a value of $\sigma_{\text{DSSSD}}=16.74(7)\,$keV. For the silicon pad detector of the Si-Si telescope a fit to the proton branch p$_{22}$, emitted from the $8975(4)\,$keV IAS in $^{21}$Na, with the Breit-Wigner width fixed to the literature value $\Gamma_{\text{tot}}=0.65(5)\,$keV gives a value of $\sigma_{\text{pad}}=22.95(14)\,$keV. For the first silicon detector of the Gas-Si-Si telescope both proton branch p$_{10}$ and p$_{22}$ were fitted and a weighted average gives $\sigma_{\text{Si1}}=20.84(14)\,$keV. The resulting FWHM values for each detector are quoted in Sect.~\ref{sec:EnergyCalib}.

Having determined the detector energy resolution the corresponding parameter $\sigma$ is fixed and the remaining parameters are fitted. In principle the detector resolution could change with the energy of the proton but this is not taken into account. This fitting procedure will lead to an overestimation of the detector resolution $\sigma$ due to the recoil broadening of the line shape but will give a more accurate determination of $\Gamma$. However, as the FWHM of the recoil broadening changes as a function of excitation energy in $^{21}$Na it will lead to a systematic bias on the fitted Breit-Wigner widths $\Gamma$. E.g. in the case of a pure Fermi $\beta$ decay the FWHM of the recoil broadening changes from 25$\,$keV at an excitation energy of 4.0$\,$MeV to 28$\,$keV at an excitation energy of 6.0$\,$MeV and to 22$\,$keV at the position of the IAS - see \cite{bib:Bha02}. Therefore a systematic uncertainty of the order of 3$\,$keV is estimated for the measured widths.

\begin{figure}
\resizebox{0.50\textwidth}{!}{
  \includegraphics{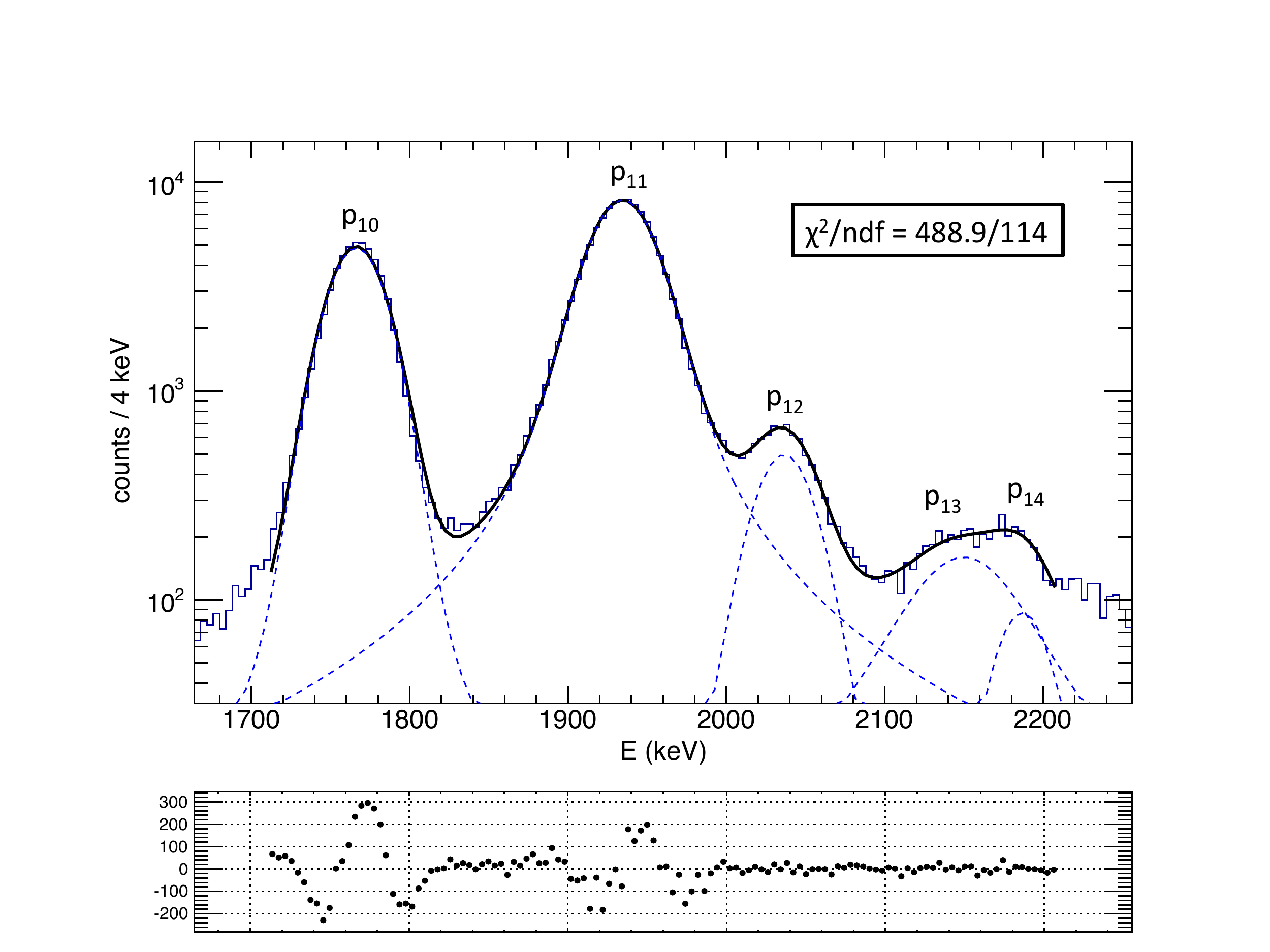}
}
\caption{(Color online) Result of a simultaneous fit of p$_{10}$, p$_{11}$, p$_{12}$, p$_{13}$, and p$_{14}$ in the DSSSD proton spectrum. Both the combined (solid black) and the individual (dashed blue) line shapes are shown - only the combined line shape is restricted to the fitting interval between 1.7 and 2.2$\,$MeV. The residuals are drawn without indicating the uncertainty.}
\label{fig:ExampleFit}
\end{figure}

We did not perform a full fit to the entire energy spectrum of one detector but divided the energy spectrum into smaller regions and assumed that the background level was zero in each region. In regions with overlapping peaks the fit included all overlapping peaks with proper normalization for each peak. An example of a fit can be seen in Figure \ref{fig:ExampleFit}. In case the proton width of a resonance is known to be smaller than the detector resolution determined, the Breit-Wigner width $\Gamma$ is fixed to the literature value.

When fitting the proton spectrum of the Gas-Si-Si telescope it was in a few cases necessary to fix the Breit-Wigner width $\Gamma$ to the value obtained in the fit of the data from the Si-Si telescope - see Table \ref{tab:Widths} for a specification of which cases (marked with '*'). Also note that in the Gas-Si-Si telescope the proton lines p$_{24}$, p$_{25}$, and p$_{26}$ are close to the end of the dynamic range of the ADC in the first silicon detector. Therefore the determined widths will be smaller than the true value as the energy calibration will no longer be linear.

It is important to note that the fitted value of $\Gamma$ will suffer from a systematic uncertainty when the detector resolution $\sigma$ is larger than the Breit-Wigner width of the resonance. If $\Gamma$ is much lower than $\sigma$ we expect the value of $\Gamma$ to be unreliable. However, as none of the resonances in $^{21}$Na have a measured width between 5--20$\,$keV we cannot tell how far below the detector resolution we can trust the determined width $\Gamma$. Hence, for proton lines where the fitted $\Gamma < \sigma$ we assign the upper limit $\Gamma \leq \sigma$. The final results for the widths are presented in Table \ref{tab:Widths}.

The quality of the fits is in general poor based on the obtained $\chi^2 /$ndf which in most cases are similar to the value obtained in the fit shown in Figure \ref{fig:ExampleFit}. However, visually the fits appear to describe the line shape well. The reason for the large $\chi^2 /$ndf is mainly a too simple response function of the detector. In a few cases an asymmetric line shape cannot be explained with the chosen line shape and this will obviously also lead to high $\chi^2$/ndf.

\begin{table}
\caption{Comparison of the fitted Breit-Wigner widths $\Gamma$ with the known literature values. For references to the literature values see Table \ref{tab:DecayScheme}. An upper limit on $\Gamma$ is set when the obtained value from the line shape fit is less than the Gaussian detector resolution. Such cases are assigned an upper limit $\Gamma \leq \sigma_{\text{det.}}$. In a few cases it were not possible to perform the line shape fit either because of low statistics, the proton line is partly punching through the detector, or it is hidden in background. Note that the fit of p$_1$ suffers from large uncertainties due to the background and the width presented here is the best estimate. The numbers marked with a * means that the DSSSD value is adopted.}
\label{tab:Widths} 
\centering
\begin{tabular}{lccc}
\hline\noalign{\smallskip}
Peak & $\Gamma_{\text{Gas-Si}}$ (keV) & $\Gamma_{\text{DSSSD}}$ (keV) & $\Gamma_{\text{tot}}$ (keV)  \\
\noalign{\smallskip}\hline\noalign{\smallskip}
p$_1$ & 65(25) & - & 21(3) \\
p$_2$ & 23(13) & $\leq$ 17 & - \\
p$_3$ & 0.65 & 0.65 & 0.65(5) \\
p$_4$ & 132(6) & 104(2) & 112(20) \\
p$_5$ & 0.01550 & 0.01550 & 0.01550(14) \\
p$_6$ & $\leq$ 21 & $\leq$ 17 & - \\
p$_7$ & 53(4) & 44(4) & 32(9) \\
p$_8$ & 0.65 & 0.65 & 0.65(5) \\
p$_9$ & 56* & 56(6) & - \\
p$_{10}$ & 3.93 & 3.93 & 3.93(10) \\
p$_{11}$ & 23.2(3) & 18.3(2) & 21(3) \\
p$_{12}$ & 26(2) & $\leq$ 17 & - \\
p$_{13}$ & 64* & 64(12)  & - \\
p$_{14}$ & 0.65 & 0.65 & 0.65(5) \\
p$_{15}$ & 136(3) & - & - \\
p$_{16}$ & 235* & 235(10) & - \\
p$_{17}$ & 133* & 133(16) & 112(20) \\
p$_{18}$ & 173(4) & 163(3) & 145(15) \\
p$_{19}$ & $\leq$ 21 & $\leq$ 23 & - \\
p$_{20}$ & $\leq$ 21 & 23(5) & 30(13) \\
p$_{21}$ & 145(7) & 145(5) & 138(16) \\
p$_{22}$ & 0.65 & 0.65 & 0.65(5) \\
p$_{23}$ & - & 204(30) & 112(20) \\
p$_{24}$ & $\leq$ 21 & 71(12) & - \\
p$_{25}$ & $\leq$ 21 & 19(4) & 30(13) \\
p$_{26}$ & - & 173(19) & 138(16) \\
p$_{27}$ & 0.65 & 0.65 & 0.65(5) \\
p$_{28}$ & - & - & 256(20) \\
\noalign{\smallskip}\hline
\end{tabular}
\end{table}

\subsubsection{Energy dependence of $\Gamma$}
\label{sec:Penetrability}
\begin{figure}
\resizebox{0.50\textwidth}{!}{
  \includegraphics{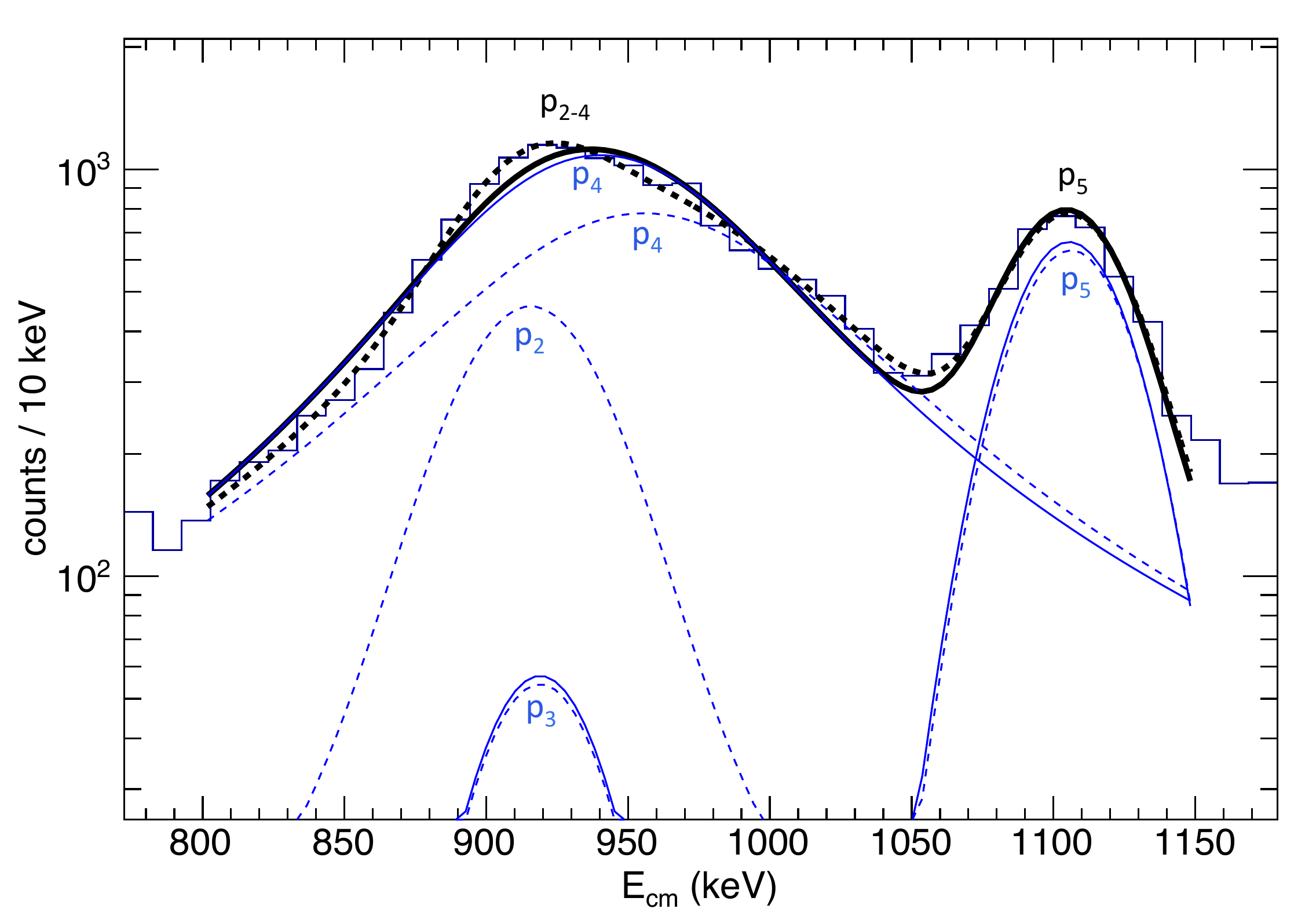}
}
\caption{(Color online) Two line shape fits with different assumptions about the contributions to the energy region are presented. The thick full drawn curve (black) assumes one broad proton branch with an energy dependent $\Gamma$ as described in the text ($L=0$ and $r_0=1.4\,$fm), the low intensity p$_3$, and p$_5$. The thin full drawn curves (blue) are the individual line shapes contributing to the total line shape. Changing the value of $L$ does not change the fit significantly. The thick dashed curve (black) assumes two proton branches, both with a constant $\Gamma$, instead of the broad proton branch. Again the low intensity p$_3$ and p$_5$ is included in the fit. The thin dashed curves (blue) are the individual line shapes contributing to the total line shape.}
\label{fig:Peak1}
\end{figure}

In the case of the combined peak called p$_{2-4}$ the line shape is asymmetric on the high energy side as is evident on Figure \ref{fig:Peak1}. The presence of the low intensity proton branch p$_3$ in this peak is known from \cite{bib:Lund}. However, it is also known to be too weak to explain the asymmetric line shape as it is seen in coincidence with the $\alpha$-particle called $\alpha_1$. Hence the proton branch p$_3$ is included in the following theoretical line shapes and the intensity of p$_3$ is fixed relative to p$_5$.
  
The observed asymmetric line shape of p$_{2-4}$ suggests either that the peak has more than two components, that interference effects with p$_5$ are present, or the width of the emitting state is large resulting in an energy dependence of the Breit-Wigner width as a result of the penetrability of the emitted charged particle through the Coulomb barrier. As the penetrability increases with the center-of-mass energy, this effect would give rise to an enhancement of the line shape towards the high energy side as observed. In the previous measurement \cite{bib:Sextro} the peak was assumed to originate from a single transition.

In order to test if the asymmetric line shape is caused by an energy dependent $\Gamma$, a refined line shape is used to fit p$_{2-4}$. By changing the fitting parameter $\Gamma$ in the following way:
\begin{equation}
\Gamma \rightarrow \Gamma (E_0)\frac{P(E,L)}{P(E_0,L)}
\end{equation} 
where $P(E,L)$ is the penetrability, $L$ is the angular momentum, and $E_0$ is the centroid energy, it is possible to model the behavior caused by the changing penetrability. The penetrability is calculated with the algorithm from \cite{bib:Penetrability} and with the nuclear radius parameter set to $r_0=1.4\,$fm. As it is evident from Figure \ref{fig:Peak1}, the chosen theoretical line shape does not reproduce the observed line shape very well (see the thick full drawn curve, $\chi^2/\text{ndf}=147.5/29=5.1$).

Another possible explanation of the observed line shape could be the presence of an additional proton branch in the peak p$_{2-4}$. Hence a theoretical line shape including two proton branches and the low intensity proton branch p$_3$ is fitted to the data. This second line shape, which is also shown in Figure \ref{fig:Peak1}, reproduces the data much better (see the thick dashed curve, $\chi^2/\text{ndf}=41.0/26=1.6$). However, the asymmetric line shape could in principle also be explained by interference. A further discussion of the interpretation of the line shape will be presented in Sect.~\ref{sec:Interpretation}.

Proton branch p$_{15}$ suffers from a similar asymmetric line shape on the high energy side. As for p$_{2-4}$ the observed line shape cannot convincingly be reproduced by including the effect of the penetrability. Hence, as for p$_{2-4}$, the line shape must be explained either by a second component in the peak or interference effects with other proton branches. A further discussion of these two options will be presented in Sect.~\ref{sec:Interpretation}.

\section{Results}
\label{results}
We present an improved interpretation of the decay scheme and we determine the absolute particle intensities and the log$(ft)$-values of the beta-decay.

\subsection{Interpretation of decay scheme}
\label{sec:Interpretation}
\begin{figure*}
\includegraphics{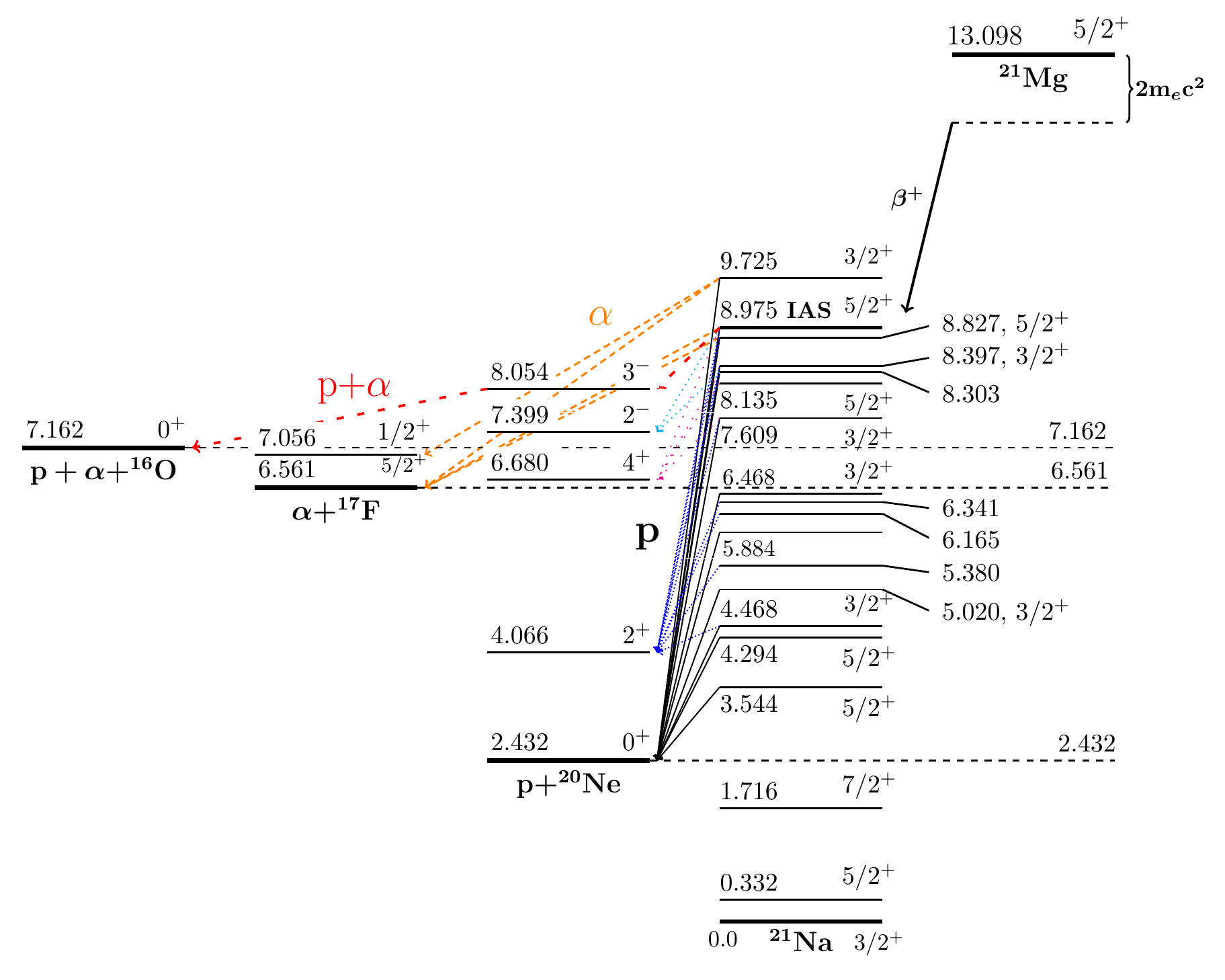}
\caption{(Color online) Decay scheme of the $^{21}$Mg $\beta$-decay. Only levels populated in the $\beta$-decay are shown. Resonances in $^{21}$Na that can be populated in allowed $\beta$-decay of $^{21}$Mg but with no definite assignment of the spin have not been assigned any spin and parity on the figure, i.e. $I^\pi=\left(\frac{3}{2}, \frac{5}{2}, \frac{7}{2} \right)^+$.}
\label{fig:DecayScheme}
\end{figure*}

A detailed interpretation of the decay scheme has been performed based on the spectroscopic properties determined by the line shape fit of the individual decay branches as described in Sect.~\ref{sec:LineShape}. For an overview of the new interpretation see Table \ref{tab:DecayScheme} and Figure \ref{fig:DecayScheme}. First, the measured energy of the charged particle together with the known levels in $^{21}$Na, which can be populated by allowed $\beta$-decay transitions, were matched up under the assumption of no new resonances in either of the daughter nuclei. Second, the measured relative particle intensities of each decay branch were compared to the known partial decay widths to establish if the assignments of ground state and excited state transitions from a given resonance in $^{21}$Na have the correct intensity ratios - several examples are given below where the total resonance width $\Gamma_{\text{tot}}$ is compared with the elastic proton scattering width $\Gamma_{\text{p,el.}}$. Third, the known total widths of the $^{21}$Na resonances $\Gamma_{\text{tot}}$ were compared with the measured widths $\Gamma$ of the individual particle decay branches obtained from the line shape fit.

As many of the known partial and total decay widths have been measured in $^{20}$Ne(p,$\,$p) scattering experiments after the work of \cite{bib:Sextro}, see \cite{bib:20Ne_proton_scattering_81,bib:20Ne_proton_scattering_92}, as well as a more recent measurement of $\gamma$p coincidences, the present interpretation has a more solid foundation.

In the process of modifying the decay scheme we have discarded 5 out of the 25 previously reported $\beta p$ branches. We also eliminate the 5979(15)$\,$keV resonance in $^{21}$Na, which together with the 5380(9)$\,$keV resonance only have been suggested in \cite{bib:Sextro}. In the new interpretation we have identified 28 different proton branches from the decay of $^{21}$Mg including seven new low intensity $\beta p$ branches and one new low intensity $\beta p\alpha$ branch (proton peak p$_3$, see \cite{bib:Lund}).

\paragraph{Asymmetry of p$_{2-4}$ and p$_{15}$:} As already mentioned in Sect. \ref{sec:Penetrability} the combined line shape of p$_{2-4}$ is asymmetric and it was shown that the line shape cannot be explained by one proton branch emitted from a broad resonance, which would give rise to an energy dependence through the penetrability. The observed line shape is much better described by replacing the one broad proton branch by two proton branches. However, the line shape may also be explained by interference effects, which would require proton emission from two close lying resonances in $^{21}$Na with the same values of $I^\pi$ and going to the same resonance in the particle daughter $^{20}$Ne. It is possible to have such interference with p$_5$ but it would mean a new interpretation of the peak p$_{2-4}$. In this case will the peak p$_{2-4}$ consist of one proton branch which interferes with p$_5$ and it is sitting on top of the low intensity p$_3$. It would, however, require to place the single proton branch and p$_5$ as proton emission from the 8303(13)$\,$keV and 8464(15)$\,$keV resonances, respectively. However, this new position of the two branches in the decay scheme, which is necessary to fulfill the conditions for interference, are unlikely as the final state would have to be the $2^-$ resonance of $^{20}$Ne. Also, it is rather unclear if p$_5$ really is asymmetric on the low energy side or if it is a distortion of the line shape caused by a partial overlap with p$_{2-4}$.

A less problematic interpretation is to assume three components in the peak p$_{2-4}$ as is demonstrated in Figure \ref{fig:Peak1}. As a consequence p$_5$ does not have to be placed as the only proton branch from the 8464(15)$\,$keV resonance of $^{21}$Na as required by the interference interpretation. Instead it can be placed much more convincingly as the ground state transition from the narrow 3544.3(4)$\,$keV resonance. Another advantage from the three component interpretation is that the relative broad proton branch p$_4$ solves the problem with the decay of the 7609(15)$\,$keV resonance. With the ground and first excited state transitions assigned to p$_{23}$ and p$_{17}$, the ratio of measured total and elastic proton scattering widths, $\Gamma_{\text{tot}}$ and $\Gamma_{\text{p,el.}}$, does not agree with the ratio of the measured intensities: $\left( \frac{I_{p,gs}}{I_{p,tot}} \right)=\left( \frac{\Gamma_{p,gs}}{\Gamma_{tot}} \right)^{exp}=0.41(7)$, which should be compared with $\left( \frac{\Gamma_{p,gs}}{\Gamma_{tot}} \right)^{ref}=0.11(3)$ from the known widths. The agreement becomes much better by including p$_4$ as a transition to the second excited state of $^{20}$Ne, $\left( \frac{I_{p,gs}}{I_{p,tot}} \right)=\left( \frac{\Gamma_{p,gs}}{\Gamma_{tot}} \right)^{exp}=0.17(2)$. The only remaining problem is the narrow proton branch p$_2$, which does not fit convincingly in the known level scheme, however, it is tentatively assigned to the 8303(13)$\,$keV resonance.

Another proton branch, p$_{15}$, was also mentioned in Sect. \ref{sec:Penetrability} as having an asymmetric line shape. As for p$_{2-4}$ the asymmetry cannot be explained by the energy dependent width. In the case of p$_{15}$ the addition of a second component does give a much better fit of the observed line shape, however, the additional proton branch does not match the known level scheme. In fact the same problem was encountered by \cite{bib:Sextro} and it is a clear sign that another explanation should be found. However, the line shape can possibly be explained by interference with p$_{11}$ which is the most intense of the $\beta$p branches from $^{21}$Mg. Both proton branches have the ground state of $^{20}$Ne as the particle daughter and as the 5020(9)$\,$keV resonance of $^{21}$Na, the parent state of p$_{15}$, has unknown $I^\pi$, interference is only possible by assigning a value of $I^\pi=\frac{3}{2}^+$ to this resonance. 

\paragraph{4467.9(7)$\,$keV resonance:} The new proton branch p$_1$ with $E_{\text{cm}}=396(3)\,$keV, which is only visible in the Gas-Si-Si telescope as the $\beta$ response of the DSSSD extends to higher energies, can only be placed at one position in the decay scheme on the basis of energy considerations. The line shape fit of p$_1$ suffers from background contamination which makes the uncertainty of the parameters large. Hence, we estimate the width of p$_1$ to be $\Gamma_{\text{DSSSD}}=65(25)\,$keV, which is in part based on the value obtained in the line shape fit.

\paragraph{5380(9)$\,$keV resonance:} Proton branch p$_6$ is interpreted as in \cite{bib:Sextro} even though the 5380(9)$\,$keV resonance in $^{21}$Na is only seen in $^{21}$Mg decay. However, in this excitation energy region $^{20}$Ne(p,p) experiments have not been performed. Looking at the measured center-of-mass energy there is basically only one other possible parent resonance for p$_6$ - the 8715(15)$\,$keV resonance with $I^\pi=3/2^+$, $\Gamma_{\text{tot}}=360(25)\,$keV, and $\Gamma_{\text{p,el.}}=160(13)\,$keV. It means that if proton branch p$_6$ were emitted from the 8715(15)$\,$keV resonance there would also have been a strong proton emission to the ground state of $^{20}$Ne with $E_{\text{cm}}=6283\,$keV - this decay branch is clearly not seen in the data (it would be 100$\,$keV below p$_{26}$). Another fact in favor of the 5380(9)$\,$keV resonance as the parent is the measurement of $\Gamma_{\text{DSSSD}}\leq 17\,$keV which rules out the very broad 8715(15)$\,$keV resonance. Finally, in \cite{bib:JCThomas} this proton line is observed in coincidence with a 1634$\,$keV $\gamma$-ray confirming the assignment of p$_6$ as a transition to the first excited state of $^{20}$Ne.

\paragraph{6165(30)$\,$keV resonance:} The proton branch p$_{12}$ with $E_{\text{cm}} = 2143.7(4)\,$keV could also be placed with the 8827(15) keV resonance based on the center-of-mass energy. But the measured width of p$_{12}$, $\Gamma_{\text{DSSSD}}\leq 17\,$keV, does not fit well with $\Gamma_{\text{tot}}=138(16)\,$keV. Hence p$_{12}$ is assigned to the 6165(30)$\,$keV resonance instead, which has an unknown total width.

\paragraph{7609(15)$\,$keV resonance:} The total width of the resonance is known to be $\Gamma_{\text{tot}}=112(20)\,$keV and the elastic proton scattering width to be $\Gamma_{\text{p,el.}}=12(3)\,$keV, see Table \ref{tab:DecayScheme}. The new proton branch p$_{23}$ with $E_{\text{cm}}=5171(7)\,$keV is placed here as a ground state transition to $^{20}$Ne. Both the energy and the width of the proton line, $\Gamma_{\text{DSSSD}}=204(30)\,$keV, are consistent with this assignment. The proton line p$_{17}$ is placed as the decay branch to the first excited state of $^{20}$Ne, which is consistent with both the energy and the width, $\Gamma_{\text{DSSSD}}=133(16)\,$keV. As the transition to the second excited state we have placed p$_4$ with $E_{\text{cm}}=936.8(1.3)\,$keV and a width of $\Gamma_{\text{DSSSD}}=104(2)\,$keV - both are consistent with the assignment. Neglecting the low intensity decay branches of $\gamma$ emission from the resonance we have the following approximate result $\left( \frac{I_{p,gs}}{I_{p,tot}} \right)=\left( \frac{\Gamma_{p,gs}}{\Gamma_{tot}} \right)^{exp}=0.17(2)$ which should be compared with $\left( \frac{\Gamma_{p,gs}}{\Gamma_{tot}} \right)^{ref}=0.11(3)$ from the known widths. The result allows for some strength to the third excited state yet to be found, however, it is expected to be a weak decay branch based on the $I^\pi=2^-$ property of the third excited state.

Note that by reinterpreting p$_{17}$ no evidence now exist for the presence of the 5979(15)$\,$keV resonance, which was introduced by \cite{bib:Sextro} to explain the proton line p$_{17}$.

\paragraph{8135(15)$\,$keV resonance:} Proton branch p$_7$ with $E_{\text{cm}}=1427.1(9)\,$keV is tentatively placed with the 8135(15)$\,$keV resonance as the parent. Based on the energy it could also match with the 8827(15)$\,$keV resonance, however, the measured width of the proton line shape agrees better with the 8135(15)$\,$keV resonance assignment as it has $\Gamma_{\text{tot}}=32(9)\,$keV. It is a problem that the ground and first excited state transitions are not observed, however, hints of a proton line below p$_{24}$ are clear, which could be explained as the missing ground state transition. But other explanations are also possible and better data with $\gamma$-proton coincidences is needed.

\paragraph{8303(13)$\,$keV resonance:} The resonance in $^{21}$Na at an energy of 8303(13)$\,$keV is only observed in the $^{23}$Na(p,t) experiment by \cite{bib:23Na_p_t} and in the $^{21}$Mg $\beta$-decay experiment by \cite{bib:Sextro}. It has tentatively been assigned $I^\pi=(\frac{3}{2}, \frac{5}{2}, \frac{7}{2})^+$ on the basis of the $\beta$-decay experiment. However, it is not observed in a $^{20}$Ne(p,p) scattering experiment \cite{bib:20Ne_proton_scattering_81}, which cover the relevant region of excitation energies in $^{21}$Na. The absence of the 8303(13)$\,$keV resonance in the scattering experiment could mean that it does not exist, however, the proton line p$_{19}$ is known from $\gamma p$ coincidences \cite{bib:JCThomas} to belong here. The measured width of p$_{19}$ is $\Gamma_{\text{DSSSD}}\leq 23\,$keV, which constrains the total width of the resonance. 

The three proton lines p$_2$, p$_9$, and p$_{24}$ can only be placed with this resonance as the parent when considering the energies. But when turning to the observed widths of the individual proton lines it is clear that an assignment of the three proton lines to the 8303(13)$\,$keV resonance can only be tentative and better data is needed to make a conclusive assignment.

\paragraph{8397(15)$\,$keV resonance:} The proton lines p$_{20}$ and p$_{25}$ are placed here as the ground and first excited state transitions to $^{20}$Ne. Based on the measured widths from the DSSSD, $\Gamma (\text{p}_{20})=23(5)\,$keV and $\Gamma (\text{p}_{25})=19(4)\,$keV, we can state that the total width of the resonance should be below the present literature value $\Gamma_{\text{tot}}=30(13)\,$keV as our measurement has a smaller uncertainty.

\paragraph{8827(15)$\,$keV resonance:} The 8827(15)$\,$keV resonance is the parent of proton branch p$_{21}$ and p$_{26}$. Neglecting the low intensity decay branches of $\alpha$ and $\gamma$ emission from the resonance we have the following approximate result $\left( \frac{I_{p,gs}}{I_{p,tot}} \right)=\left( \frac{\Gamma_{p,gs}}{\Gamma_{tot}} \right)^{exp}=0.21(2)$ to be compared with $\left( \frac{\Gamma_{p,gs}}{\Gamma_{tot}} \right)^{ref}=0.28(5)$. The two numbers agree within one standard deviation and as the decay branch $\alpha_2$, which also goes through this resonance, is tiny the assignment of p$_{21}$ and p$_{26}$ to the 8827(15)$\,$keV resonance is consistent.

\begin{table}
\caption{$E_{\text{cm}}$ in keV and relative proton and $\alpha$ particle intensities with respect to p$_{11}$ compared to earlier results \cite{bib:Sextro,bib:JCThomas}. We adopted the values from the Si-Si telescope due to less $\beta$-summing and better background reduction by the use of the DSSSD. However, for p$_1$, p$_{15}$, and p$_{28}$ the value from the Gas-Si-Si telescope is adopted. Note that the relative intensity of p$_3$ and $\alpha_1$ is the same as they are emitted in coincidence as a $\beta p\alpha$ decay branch. For more details on the $\beta\alpha$ and $\beta p\alpha$ emissions see \cite{bib:Lund}. \#: p$_{13}$--p$_{14}$ and p$_2$--p$_4$ were observed as a single proton line in \cite{bib:Sextro,bib:JCThomas}.}
\label{tab:ProtonIntensity} 
\centering
\begin{tabular}{ccccc}
\hline\noalign{\smallskip}
 & $E_{\text{cm}}$ (keV) & \multicolumn{3}{c}{Relative intensity (\%)} \\
\noalign{\smallskip}\hline\noalign{\smallskip}
 & This work & This work & \cite{bib:Sextro} & \cite{bib:JCThomas} \\
\noalign{\smallskip}\hline\noalign{\smallskip}
p$_1$ & 396(3) & 3.91(45) & - & - \\
p$_2$ & 906.2(14) & 2.0(5) & \#16.6(6) & \#8.3(35) \\
p$_3$ & 919(18) & 0.28(3) & \#16.6(6) & \#8.3(35) \\
p$_4$ & 936.8(13) & 19.4(5) & \#16.6(6) & \#8.3(35) \\
p$_5$ & 1101.8(3) & 3.34(6) & 4.3(6) & 8.6(23) \\
p$_6$ & 1315.7(1) & 20.01(15) & 23.3(17) & 18.1(33) \\
p$_7$ & 1427.1(9) & 2.84(11) & - & -  \\
p$_8$ & 1564.0(3) & 4.66(9) & 6.3(6) & - \\
p$_9$ & 1630.0(15) & 2.95(17) & - & - \\
p$_{10}$ & 1860.62(8) & 44.05(24) & 51.4(21) & 46.1(73) \\
p$_{11}$ & 2036.75(8) & 100.0(4) & 100.0(15) & 100(15) \\
p$_{12}$ & 2143.7(4) & 4.58(14) & 8.3(10) & - \\
p$_{13}$ & 2263(4) & 3.79(55) & \#4.6(4) & - \\
p$_{14}$ & 2302(2) & 0.73(20) & \#4.6(4) & - \\
p$_{15}$ & 2587.2(14) & 20.89(24) & 7.6(20) & 7.0(21) \\
- & - & - & 2.9(10) & - \\
- & - & - & 3.3(2) & - \\
p$_{16}$ & 3443(2) & 34.6(31) & 5.9(3) & $<$21.5(35) \\
p$_{17}$ & 3585(4) & 8.0(15) & 4.5(5) & $<$11.8(22) \\
- & - & - & 1.3(1) & - \\
- & - & - & 1.3(2) & - \\
- & - & - & 3.6(5) & - \\
p$_{18}$ & 4054.5(10) & 33.58(2.45) & 10.3(4) & 9.3(19) \\
p$_{19}$ & 4256.7(12) & 1.99(20) & 2.5(2) & 4.4(13) \\
p$_{20}$ & 4356.0(16) & 1.94(19) & 1.0(2) & - \\
p$_{21}$ & 4769(2) & 10.9(8) & 2.5(5) & - \\
p$_{22}$ & 4912.5(2) & 24.29(176) & 14.9(7) & 24.7(43) \\
p$_{23}$ & 5171(7) & 5.63(75) & - & - \\
p$_{24}$ & 5868(3) & 1.56(18) & 0.5(6) & - \\
p$_{25}$ & 5983.2(13) & 1.37(13) & 0.7(6) & - \\
p$_{26}$ & 6388(5) & 2.86(29) & 0.6(9) & - \\
p$_{27}$ & 6536.9(3) & 8.85(65) & 5.5(4) & 9.7(18) \\
p$_{28}$ & 7200(300) & 0.05(2) & - & - \\
$\alpha_1$ & 882(15) & 0.28(3) & - & - \\
$\alpha_2$ & 2201(25) & 0.11(1) & - & - \\
$\alpha_3$ & 2397(3) & 1.79(5) & - & - \\
$\alpha_4$ & 2700(42) & 0.10(1) & - & - \\
$\alpha_5$ & 3060(80) & 0.04(1) & - & - \\
\noalign{\smallskip}\hline
\end{tabular}
\end{table}

\paragraph{8975(4)$\,$keV resonance (IAS):} As expected several proton branches originate from the IAS: p$_3$, p$_8$, p$_{14}$, p$_{22}$, and p$_{27}$. Including the $\alpha$ emission but neglecting the low intensity decay branches of $\gamma$ emission from the resonance, we have the following approximate result $\left( \frac{I_{p,gs}}{I_{p,tot}+I_{\alpha ,tot}} \right)=\left( \frac{\Gamma_{p,gs}}{\Gamma_{tot}} \right)^{exp}=0.22(2)$ to be compared with $\left( \frac{\Gamma_{p,gs}}{\Gamma_{tot}} \right)^{ref}=0.18(2)$. The two numbers are consistent with each other and the measured $\beta$-decay strength to the IAS is consistent with the sum rule ($\Sigma B_F =Z-N$) within one standard deviation. Also note that the assignment of p$_{22}$ with the IAS has been confirmed by the measurement of $\gamma p$ coincidences \cite{bib:JCThomas}.

\paragraph{9725(25)$\,$keV resonance:} As the energy calibration of the second silicon detector in the Gas-Si-Si telescope suffers from large uncertainties it makes it difficult to place p$_{28}$ in the decay scheme. However, it is certain that the $^{21}$Na resonance emitting the proton is located above the IAS as p$_{27}$ is a transition between the IAS and the ground state of $^{20}$Ne - see Figure \ref{fig:ProtonSpectrumSi1}. Based on the $^{20}$Ne(p,p) scattering experiment in Ref. \cite{bib:20Ne_proton_scattering_81} three resonances above the IAS are available for allowed $\beta$-decay of $^{21}$Mg and are therefore possible candidates: 8981(15)$\,$keV, 9155(15)$\,$keV, and 9725(25)$\,$keV. We favor the 9725(25)$\,$keV resonance as the parent of p$_{28}$ as the $\beta\alpha$ branch $\alpha_5$ has to go through this resonance based on the measured $E_{\text{cm}}$. A further argument is that also $\alpha_4$ is most likely emitted from this resonance. Even though this resonance emits $\alpha$-particles a calculation of the reduced widths, $\gamma^2_c / \gamma_{sp}^2$, shows no evidence for $\alpha$-clustering.

In the proton spectrum of the Si-Si telescope, see Figure \ref{fig:ProtonSpectrumDSSSD}, the proton line p$_{28}$ is tentatively placed at the expected energy. A similar structure to the one observed at energies above the peak p$_{27}$, where the proton branch p$_{28}$ is observed, is seen in the energy region between p$_{22}$ and p$_{24}$. Studying closer the energy region just below p$_{24}$ hints of a weak, but broad proton line are observed at a center-of-mass energy of 5600--5750$\,$keV. Together with considerations on the intensity it can be assigned as proton emission from the 9725(25)$\,$keV resonance to the first excited state of $^{20}$Ne. Alternatively it can be assigned as the ground state transition from the 8135(15)$\,$keV resonance. As the 9725(25)$\,$keV resonance is broad, $\Gamma_{\text{tot}}=256(20)\,$keV, the energy dependence of the $\beta$-phase space factor may explain both the structure between p$_{22}$ and p$_{24}$ and the structure above p$_{27}$. However, the present data is not sufficient to confirm this hypothesis. Future experiments should hopefully be able to solve this issue. 

The improved decay scheme proposed for the $^{21}$Mg $\beta$-decay is shown in Figure \ref{fig:DecayScheme} and in Table \ref{tab:DecayScheme} the corresponding known total widths and the known elastic proton scattering widths are given. All of the ambiguities in the interpretation of the decay scheme could in principle be solved by a dedicated measurement of the decay with both $\gamma$-ray and charged particle detection. Any coincidence spectra of charged particles and $\gamma$-rays will give a clear signature of the resonances responsible for the decay branch. The most intense $\beta p$ transitions to excited states in $^{20}$Ne have already been measured in coincidence with the 1634$\,$keV $\gamma$-ray from the de-excitation of the first excited state in $^{20}$Ne \cite{bib:JCThomas}. However, to observe $\gamma p$ coincidences from the less intense transitions considerably more statistics is required.

\begin{table*}
\caption{Left part of the table compares the weighted average of the $^{21}$Na resonances with the literature energies. Previous knowledge about the $^{21}$Na resonances are also presented. Here $I^{\pi}$ is the spin and parity, $\Gamma_{\text{tot}}$ is the total width of the resonance, and $\Gamma_{\text{p,el}}$ is the elastic proton scattering width of the resonance. If $I^\pi=\left( \frac{3}{2}, \frac{5}{2}, \frac{7}{2} \right)^+$ it is marked by a $\#$. The right part of the table shows the new interpretation of the decay scheme. If the proton line is in italic, \textit{p$_i$}, the assignment is not conclusive. An x marks an unobserved but energetically allowed proton emission. The particle emission thresholds are S$_p$($^{21}$Na)$=2431.68(28)\,$keV and S$_\alpha$($^{21}$Na)$=6561.3(4)\,$keV, \cite{bib:Nubase}. The quoted uncertainties on the weighted average of the $^{21}$Na resonances, $\bar{E}^*_{\text{meas.}}$, do not take the calibration uncertainty into account - only the statistical uncertainty from the fit.}
\label{tab:DecayScheme} 
\resizebox{\textwidth}{!}{
\begin{tabular}{ccccccccccc}
\hline\noalign{\smallskip}
$E^*(^{21}$Na) & $\bar{E}^*_{\text{meas.}}$ (keV) & I$^\pi$ & $\Gamma_{\text{tot}}$ (keV) & $\Gamma_{\text{p,el.}}$ (keV) & Reference & \multicolumn{5}{c}{$^{20}$Ne resonances (MeV, $I^\pi$)} \\
\noalign{\smallskip}\hline\noalign{\smallskip}
  & & & & & & 0.0, $0^+$ & 1.63, $2^+$ & 4.25, $4^+$ & 4.97, $2^-$ & 5.62, $3^-$ \\ 
\noalign{\smallskip}\hline\noalign{\smallskip}
0.0, T$=\frac{1}{2}$ & & $3/2^+$ & - & - &  &  & & & & \\
331.90(10) & & $5/2^+$ & - & - &  & &  & & \\
1716.1(3) & & $7/2^+$ & - & - &  &  & & & \\
3544.3(4) & 3533.5(4) & $5/2^+$ & 0.01550(14) & - & \cite{bib:20Ne_proton_scattering_69} & p$_5$ & & & & \\
4294.3(6) & 4292.3(3) & $5/2^+$ & 3.93(10) & - & \cite{bib:20Ne_proton_scattering_69} & p$_{10}$ & x & & & \\
4467.9(7) & 4468.4(3) & $3/2^+$ & 21(3) & Observed & \cite{bib:20Ne_proton_scattering_64} & p$_{11}$ & p$_1$ & & & \\
5020(9) & 5018.9(14) & $3/2^+$ & - & - &  & p$_{15}$ & x & & & \\
5380(9) & 5381.1(3) & $\#$ & - & - &  & x & p$_6$ & & & \\
5770(20) & - & $\#$ & $\sim$20 & - & \cite{bib:Firestone}  & x & x & & & \\
5884(20) & 5874.8(24) & $\#$ & - & - &  & p$_{16}$ & x & & & \\
5979(15) & - & $\#$ & - & - &  & x & x & & & \\
6165(30) & 6209.1(5) & $\#$ & - & - &  & x & p$_{12}$ & & & \\
6341(20) & 6328(4) & $\#$ & - & - &  & x & p$_{13}$ & & & \\
6468(20) & 6486.2(10) & $3/2^+$ & 145(15) & 130(12) & \cite{bib:20Ne_proton_scattering_81}  & p$_{18}$ & x & & & \\
7609(15) & 7620(2) & $3/2^+$ & 112(20) & 12(3) & \cite{bib:20Ne_proton_scattering_81}  & p$_{23}$ & p$_{17}$ & p$_{4}$ & x & \\
8135(15) & 8106.5(13) & $5/2^+$ & 3(9) & 7(3) & \cite{bib:20Ne_proton_scattering_81}  & x & x & \textit{p$_7$} & x & x \\
8303(13) & 8312.8(8) & $\#$ & - & - &  & \textit{p$_{24}$} & p$_{19}$ & \textit{p$_9$} & \textit{p$_{2}$} & x \\
8397(15) & 8417.5(10) & $3/2^+$ & 30(13) & 5(2) & \cite{bib:20Ne_proton_scattering_81}  & p$_{25}$ & p$_{20}$ & x & x & x \\
8464(15) & - & $3/2^+$ & 25(9) & 5(1) & \cite{bib:20Ne_proton_scattering_81}  & x & x & x & x & x \\
8562(15) & - & $3/2^+$ & $<20$ & 1.0(5) & \cite{bib:20Ne_proton_scattering_81}  & x & x & x & x & x \\
8595(15) & - & $5/2^+$ & 138(15) & 25(5) & \cite{bib:20Ne_proton_scattering_81}  & x & x & x & x & x \\
8715(15) & - & $3/2^+$ & 360(25) & 160(13) & \cite{bib:20Ne_proton_scattering_81}  & x & x & x & x & x \\
8827(15) & 8833(2) & $5/2^+$ & 138(16) & 38(5) & \cite{bib:20Ne_proton_scattering_81}  & p$_{26}$ & p$_{21}$ & x & x & x \\
8975(4), T$=\frac{3}{2}$ & 8971.1(2) & $5/2^+$ & 0.65(5) & 0.117(10) & \cite{bib:20Ne_proton_scattering_92}  & p$_{27}$ & p$_{22}$ & p$_{14}$ & p$_8$ & p$_3$ \\
8981(15) & - & $5/2^+$ & 23(16) & 2.5(1.0) & \cite{bib:20Ne_proton_scattering_81}  & x & x & x & x & x \\
9155(15) & - & $3/2^+$ & 34(13) & 8(2) & \cite{bib:20Ne_proton_scattering_81}  & x & x & x & x & x \\
9280(30) & - & $\#$ & - & - &  & x & x & x & x & x \\
9725(25) & 9632(300) & $3/2^+$ & 256(20) & 136(15) & \cite{bib:20Ne_proton_scattering_81}  & p$_{28}$ & x & x & x & x \\
\noalign{\smallskip}\hline
\end{tabular}
}
\end{table*}

\subsection{Relative proton intensities}
Based on the line shape fit of the proton spectrum the relative branching ratios have been determined. The results can be seen in Table \ref{tab:ProtonIntensity} including a comparison with earlier works \cite{bib:Sextro,bib:JCThomas}. The new values are expected to be more reliable than the old values because of the use of a more reliable line shape, which accounts for the physics better than that previously used. Also the relative proton intensities are based on the data obtained with the Si-Si telescope, which has several advantages. First, the DSSSD has a better energy resolution than the Gas-Si-Si telescope. Second, the amount of $\beta$-summing is minimal due to the small solid angle of each of the 256 pixels of the DSSSD, $\Omega_{\text{pixel}}=0.030(1)\,\%$. Third, the data acquisition threshold on the gas detector is cutting part of the high energy protons - cf. Figure \ref{fig:BananaPlot}.

To obtain consistent relative proton intensities from the Si-Si telescope it is important to be aware of the different solid angles for the two detectors. To remedy this a scaling factor has been applied to the number of protons extracted from the back detector which is just the ratio of the solid angles: $\frac{\Omega_{\text{dsssd}}}{\Omega_{\text{back}}}=1.17(8)$.

\subsection{Absolute intensities and $\mathbf{\beta}$ branching ratios}
\label{sec:logft}
The measured relative proton intensities as quoted in Table \ref{tab:ProtonIntensity} follow the same systematic trend with respect to \cite{bib:Sextro} as the data of \cite{bib:JCThomas}, i.e. for p$_1$ to p$_{14}$ the relative proton intensities tend to be systematically lower while for p$_{15}$ to p$_{28}$ the relative intensities tend to be systematically larger. This indicates that the experiment of \cite{bib:Sextro} is suffering from a systematic error. Hence we use the results of \cite{bib:JCThomas} to scale the measured relative proton intensities to get the absolute proton intensities. From the ratio of the measured absolute and relative intensities of proton branch p$_{22}$ and p$_{28}$ we get the relevant scaling factor to be $R=0.057(9)$.

From the four observed $\beta\alpha$ branches we extract the relative $\alpha$-particle intensities with respect to the proton branch p$_{11}$ in the same way as for the relative proton intensities. Then the absolute $\alpha$-particle intensities are determined by use of the scaling factor $R$ in the same way as it was done for the protons.

The $\beta$-decay branching ratios are then calculated directly from the absolute proton and $\alpha$-particle intensities as the $\gamma$-decay widths are of the order $10^2$ to $10^3$ times smaller than the charged particle decay widths. Hence we do not correct the absolute intensities for the possible $\gamma$-decays in converting to $\beta$-decay branching ratios. The resulting log$(ft)$-values are presented in Table \ref{tab:BranchingRatio}. The log$(ft)$-values are calculated according to the parametrization of \cite{bib:logft} with $Q_{\beta^+}=12.076(16)\,$MeV. The uncertainty on the log$(ft)$-value is dominated by uncertainty on the absolute $\beta$-decay branching ratio. The relative uncertainty on the absolute $\beta$-decay branching ratios are an order of magnitude larger than the relative uncertainty on both the half-life and the excitation energy in $^{21}$Na.

Note that the measured log$(ft)$-values for the $5884(20)\,$ keV and the $8303(13)\,$keV resonances, see Table \ref{tab:BranchingRatio}, are so low that the $\beta$-decay to these two resonances definitely is allowed, i.e. the two resonances must have $I^\pi =\left( \frac{3}{2},\frac{5}{2},\frac{7}{2}\right)^+$.

\begin{table*}
\caption{Absolute $\beta$ branching ratios and log(ft)-values compared to earlier results by \cite{bib:Sextro,bib:JCThomas}.}
\label{tab:BranchingRatio} 
\centering
\begin{tabular}{cccccccccc}
\hline\noalign{\smallskip}
$E^*(^{21}$Na) (keV) & \multicolumn{3}{c}{\text{B.R. from $^{21}$Mg} (\%)} & \multicolumn{3}{c}{\text{log($ft$)}} \\
\noalign{\smallskip}\hline\noalign{\smallskip}
 & This work & \cite{bib:Sextro} & \cite{bib:JCThomas}& This work & \cite{bib:Sextro} & \cite{bib:JCThomas}  \\
\noalign{\smallskip}\hline\noalign{\smallskip}
3544.3(4) & 0.19(3) & 0.45(7) & 0.7(2) & 6.45(16) & 6.09(6) & 5.90(17) \\
4294.3(6) & 2.5(4) & 5.36(31) & 3.7(6) & 5.14(16) & 4.82(2) & 4.99(8) \\
4467.9(7) & 5.89(96) & 10.45(46) & 8.00(13) & 4.73(16) & 4.48(2) & 4.61(8) \\
5020(9) & 1.18(19) & 2.53(25) & 1.2(3) & 5.27(16) & 4.95(4) & 5.26(16) \\
5380(9) & 1.13(18) & 2.43(21) & 1.5(3) & 5.18(16) & 4.85(3) & 5.08(10) \\
5770(20) & 0.0 & 0.34(3) & 0.0 & - & 5.59(3) & - \\
5884(20) & 1.96(36) & 0.62(4) & $<$1.7(3) & 4.90(18) & 5.30(3) & $>$4.87(10) \\
5979(15) & 0.0 & 0.47(6) & $<$0.9(2) & - & 5.38(5) & $>$5.07(10) \\
6094(35) & 0.0 & 0.14(1) & 0.0 & - & 5.86(3) & - \\
6165(30) & 0.26(4) & 0.0 & 0.0 & 5.68(15) & - & - \\
6210(50) & 0.0 & 0.14(2) & 0.0 & - & 5.84(6) & - \\
6341(20) & 0.21(5) & 0.86(7) & 0.0 & 5.71(24) & 4.99(3) & - \\
6468(20) & 1.90(34) & 1.07(6) & 0.8(2) & 4.71(18) & 4.84(2) & 5.00(11) \\
7609(15) & 1.88(36) & 0.0 & 0.0 & 4.26(19) & - & - \\
8135(15) & 0.16(3) & 0.0 & 0.0 & 5.08(19) & - & - \\
8303(13) & 0.5(1) & 0.31(3) & 0.4(1) & 4.5(2) & 4.60(3) & 4.56(20) \\
8397(15) & 0.19(4) & 0.0 & 0.0 & 4.87(21) & - & - \\
8464(15) & 0.0 & 0.18(2) & 0.0 & - & 4.77(5) & - \\
8827(15) & 0.79(14) & 1.19(13) & 0.0 & 4.01(18) & 3.73(4) & - \\
8975(4) & 2.3(4) & 2.79(16) & 2.8(4) & 3.45(17) & 3.26 & 3.27(8) \\
9725(25) & 0.010(2) & 0.0 & 0.0 & 5.3(2) & - & - \\
\noalign{\smallskip}\hline
\end{tabular}
\end{table*}

\section{Discussion}
\label{discussion}
We start with a discussion of systematic effects and uncertainties. In the case of the measured Breit-Wigner widths $\Gamma$, which were presented in Table \ref{tab:Widths}, the values obtained from the first silicon detector of the Gas-Si-Si telescope for the proton branches p$_{24}$, p$_{25}$, and p$_{26}$ all suffer from a systematic bias. This is caused by the fact that the assumed linear energy calibration is no longer valid close to the end of the ADC range. It results in systematically smaller center-of-mass energies and Breit-Wigner widths.

Another point is the uncertainty associated with the energy calibration. The quoted uncertainties on the determined center-of-mass energies, which are presented in Table \ref{tab:ProtonIntensity}, only account for the statistical uncertainty of the line shape fit. However, the energy calibrations do suffer from systematic uncertainties which will add to the statistical uncertainty of the line shape fit. From the proton branches emitted from resonances in $^{21}$Na which also decay by $\gamma$ emission the systematic uncertainty on the energy calibration is estimated to be close to 10$\,$keV.

\subsection{Relative proton intensity}
The relative proton intensities measured were compared with earlier measurements by \cite{bib:Sextro,bib:JCThomas} in Table \ref{tab:ProtonIntensity}. On several points we disagree with either one or both of the earlier measurements, but the main trend is the same. As already mentioned a systematic deviation with respect to the results of \cite{bib:Sextro} is observed. For p$_1$ to p$_{14}$ the relative proton intensities tend to be systematic lower than measured by \cite{bib:Sextro}, while for p$_{15}$ to p$_{28}$ the relative intensities tend to be systematically larger. The results of \cite{bib:JCThomas} appear to follow the same systematic trend but some differences with our results remain. It does, however, indicate a systematic error in the results of \cite{bib:Sextro}.

A closer look on the methods used in the earlier experiments makes it clear that the results, which we have obtained in the present work are more reliable. Both earlier measurements suffer from $\beta$-summing effects due to the use of non-segmented detectors. The solid angle of the 256 individual pixels of the DSSSD is on average $0.030(1)\,$\% and it makes $\beta$-summing effects negligible. In addition the line shape function used here constitutes an improvement compared to the normalized Gaussian functions used in \cite{bib:Sextro,bib:JCThomas}, which help to minimize systematic errors due to a line shape that does not describe the physics case adequately.

\section{Summary and conclusions}
\label{conclusion}
The $\beta$-delayed emission of charged particles from $^{21}$Mg has been measured at ISOLDE, CERN, with a set of charged particle telescopes. This has resulted in several new results which is summarized in the following list.

\begin{itemize}
\item Seven new $\beta p$ branches were observed from the decay of $^{21}$Mg out of a total of 27 $\beta p$ branches measured with energies between 0.39$\,$MeV and 7.2$\,$MeV.
\item An improved interpretation of the decay of $^{21}$Mg has been proposed on the basis of the extracted spectroscopic information obtained with a fit of the proton line shape. It is consistent with the results obtained in reaction studies.
\item As a consequence of the interpretation of the line shape of p$_{15}$ to be due to the effect of interference with p$_{11}$, an assignment of $I^\pi=\frac{3}{2}^+$ for the 5020(9)$\,$keV resonance in $^{21}$Na is made. 
\item The peak p$_{17}$ has been assigned to the feeding to the first excited state, so the need of a resonance at 5979(15)$\,$ keV only proposed in \cite{bib:Sextro} is questionable.
\item The total width $\Gamma_{\text{tot}}$ of the 5020(9)$\,$keV, 5380(9)$\,$keV, 5884(20)$\,$keV, 6165(30)$\,$keV, 6341(20)$\,$keV, 8135(15) keV, and 8303(13)$\,$keV resonances in $^{21}$Na has been measured for the first time here - see Table \ref{tab:Widths} and \ref{tab:DecayScheme}.
\item For the 6468(20)$\,$keV resonance in $^{21}$Na we constrain the total width to be close to the upper uncertainty limit of the previously measured value: $\Gamma_{\text{tot}}=145(15)$ keV.
\item For the 8397(15)$\,$keV resonance in $^{21}$Na we constrain the total width to be close to the lower uncertainty limit of the previously measured value: $\Gamma_{\text{tot}}=30(13)$ keV.
\item Finally we measured the half-life of $^{21}$Mg to be $T_{1/2}=118.6(5)\,$ms which is a factor of four improvement on the statistical uncertainty compared to the literature value.
\end{itemize}
The modified interpretation calls for a dedicated measurement of both charged particles and $\gamma$-rays to verify the new interpretation. The detection of the $\gamma$-rays would also make a precise and accurate determination of absolute proton intensities possible.  
\newline
\newline
This work has been supported by the European Commision within the Seventh Framework Programme "European Nuclear Science and Applications Research", contract no. 262010 (ENSAR), and by the Spanish research agency under number FPA2012-32443.


\begin{thebibliography}{}
\bibitem{bib:Borge08}
B. Blank and M. J. G. Borge, Prog. Part. Nucl. Phys. \textbf{60}, 403 (2008).
\bibitem{bib:Pfutzner}
M. Pf\"{u}tzner, L.V. Grigorenko, M. Karny, and K. Riisager, Rev. Mod. Phys. \textbf{84}, 567 (2012).
\bibitem{bib:Borge13}
M. J. G. Borge, Phys. Scr. T\textbf{152}, 014013 (2013).
\bibitem{bib:Nubase} 
G. Audi et al., Chin. Phys. C {\bf 36}, 1157 (2012).
\bibitem{bib:Sextro}
Richard G. Sextro, R. A. Gough, and Joseph Cerny, Phys. Rev. C \textbf{8}, 258 (1973).
\bibitem{bib:JCThomas}
J.-C. Thomas PhD thesis, University of Bordeaux, 2003.
\bibitem{bib:isol}
E. Kugler, Hyperfine Interact. \textbf{129}, 23 (2000).
\bibitem{bib:Rilis}
V.N. Fedoseyev, G. Huber, U. K\"{o}ster, J. Lettry, V.I. Mishin, H. Ravn, V. Sebastian, Hyperfine Interact. \textbf{129}, 409 (2000).
\bibitem{bib:BetaSumming}
D. Schardt and K. Riisager, Z. Phys. A \textbf{345}, 265-271 (1993).
\bibitem{bib:Lennard}
W. N. Lennard, H. Geissel, K. B. Winterbon, D. Phillips, T. K. Alexander, and J. S. Forster, Nucl. Inst. Meth. in Phys. Res. A \textbf{248} 454-460 (1986).
\bibitem{bib:SRIM}
J. F. Ziegler, J. P. Biersack, and M. D. Ziegler, \textit{SRIM - The Stopping and Range of Ions in Matter}, 5th ed. (SRIM Co., USA, 2008).
\bibitem{bib:Lund}
M.V. Lund, M.J.G. Borge, J.A. Briz, J. Cederk\"{a}ll, H.O.U. Fynbo, J.H. Jensen, B. Jonson, K.L. Laursen, T. Nilsson, A. Perea, V. Pesudo, K. Riisager, and O. Tengblad, arXiv:1506.03915v1 (2015).
\bibitem{bib:Bergmann}
U. C. Bergmann and K. Riisager, Nucl. Phys. A \textbf{701}, 213c (2002).
\bibitem{bib:Ste86} 
M.A. Stephens, p. 97 in \textit{Goodness-of-fit techniques}, eds R.B. D'Agostino and M.A. Stephens (Marcel Dekker, New York, 1986).
\bibitem{bib:Cho94} 
V. Choulakian, R.A. Lockhart and M.A. Stephens, Canad. J. Statist. {\bf 22}, 125 (1994).
\bibitem{bib:Rmatrix}
A. M. Lane and R. G. Thomas, Rev. Mod. Phys. \textbf{30}, (1958).
\bibitem{bib:Bha02}  
M. Bhattacharya and E. G. Adelberger, Phys. Rev. C, \textbf{65}, 055502 (2002).
\bibitem{bib:Penetrability}
N. Michel, Comp. Phys. Comm. \textbf{176}, 232 (2007).
\bibitem{bib:20Ne_proton_scattering_81}
M. Fern\'{a}ndez, G. Murillo, J. Ramirez, O. Avila, S.E. Darden, M.C. Rozak, J.L. Foster, B.P. Hichwa, and P.L. Jolivette, Nucl. Phys. A \textbf{369}, 425 (1981).
\bibitem{bib:20Ne_proton_scattering_92}
J.F. Wilkerson, T.M. Mooney, R.E. Fauber, T.B. Clegg, H.J. Karwowski, E.J. Ludwig, and W.J. Thompson, Nucl. Phys. A \textbf{549}, 223 (1992).
\bibitem{bib:23Na_p_t}
G. W. Butler, J. Cerny, S. W. Cosper, and R. L. McGrath, Phys. Rev. \textbf{166}, 1096 (1968).
\bibitem{bib:Firestone}
R. B. Firestone,	Nuclear Data Sheets \textbf{103}, 269 (2004).
\bibitem{bib:logft}
D. H. Wilkinson and B. E. F. Macefield, Nucl. Phys. A \textbf{232}, 58 (1974).
\bibitem{bib:20Ne_proton_scattering_64}
C. Van der Leun and W.L. Mouton, Physica \textbf{30}, 333 (1964).
\bibitem{bib:20Ne_proton_scattering_69}
R. Bloch, T. Knellwolf, and R.E. Pixley, Nucl. Phys. A \textbf{123}, 129 (1969).
\end{thebibliography}
\end{document}